\documentclass[a4paper,11pt]{article}
\pdfoutput=1 
\usepackage{jheppub}

\usepackage[font=small,labelfont=bf]{caption}
\usepackage[T1]{fontenc}
\usepackage[utf8]{inputenc}
\usepackage[normalem]{ulem}

\usepackage{amsmath}
\usepackage{amssymb}
\usepackage{amsfonts}
\usepackage{empheq} 
\usepackage{physics}
\usepackage{blkarray}

\usepackage{graphicx}
\usepackage{subcaption}
\usepackage{tikz}
\usepackage[most]{tcolorbox} 

\DeclareMathOperator{\diag}{diag}

\usepackage{placeins} 

\usepackage[colorlinks=true,linkcolor=blue,citecolor=blue, urlcolor=blue]{hyperref} 

\graphicspath{{../Images/}}

\title{Adiabatic hydrodynamization with transverse spatial gradients in boost-invariant plasmas}

\author[a]{Uri Sharell,}
\author{Jasmine Brewer,}
\author[b,c]{and Weiyao Ke}

\affiliation[a]{Institut für Theoretische Physik, Universität Heidelberg, 69120 Heidelberg, Germany}
\affiliation[b]{Key Laboratory of Quark and Lepton Physics (MOE) \& Institute of Particle Physics, Central China Normal University, Wuhan 430079, China}
\affiliation[c]{Southern Center for Nuclear-Science Theory (SCNT), Institute of Modern Physics, Chinese Academy of Sciences, Huizhou 516000, China}

\emailAdd{sharell@thphys.uni-heidelberg.de}
\emailAdd{weiyaoke@ccnu.edu.cn}

\abstract{The quantitative success of relativistic viscous hydrodynamics in describing the short-lived quark--gluon plasma raises a fundamental question: how does far-from-equilibrium QCD matter approach hydrodynamic behavior so rapidly? Microscopic kinetic-theory studies have related this onset to attractor dynamics, but typically assume transverse homogeneity. We relax this assumption by introducing gradient modes with finite transverse wave number $k$, extending the analysis of Ref.~\cite{brewerFarfromequilibriumSlowModes2022}. These couple different spherical harmonics of the momentum distribution, and the resulting dynamics is controlled by the competition among the expansion rate $1/\tau$, the collision rate $1/\tau_R$, and the gradient scale $k$. At early times, longitudinal expansion suppresses this coupling, and different spherical harmonic sectors follow their homogeneous attractors. We find that, at later times and sufficiently small $k$, this gradient coupling drives the system toward a global attractor manifold spanned by hydrodynamic sound and shear modes, on a timescale $\tau_D$ that depends on both the azimuthal harmonic $m$ and $k\tau_R$. For sufficiently large $k\tau_R$, the spectral gap closes: perturbations retain finite damping rates and therefore equilibrate, but no longer follow an isolated hydrodynamic attractor or admit a reduced description involving only a few hydrodynamic modes.
}
\begin{document}
\maketitle

\newpage

\section{Introduction}\label{sec:Intro}
The discovery of deconfined QCD matter, the quark--gluon plasma (QGP), in ultrarelativistic heavy-ion collisions at RHIC~\cite{BRAHMS:2004adc} and the LHC~\cite{Muller:2012zq} has ushered in a new era in the study of equilibration in QCD. One of the most striking findings is the strong collective flow exhibited by the produced matter~\cite{ALICE:2010suc}. A broad range of bulk observables, including transverse-momentum spectra and anisotropic-flow coefficients, can be quantitatively described by relativistic viscous hydrodynamics~\cite{Heinz:2013th,Gale:2013da}.
Such calculations based on hydrodynamics have shown impressive agreement with experimental measurements~\cite{Romatschke:2007mq,Heinz:2005bw,Gale:2012rq} and have made it possible to constrain transport coefficients of the quark--gluon plasma~\cite{Novak:2013bqa,Bernhard:2015hxa,Bernhard:2016tnd,Bernhard:2019bmu,Niemi:2015qia,JETSCAPE:2020mzn,JETSCAPE:2020shq}. A major conclusion of phenomenological analyses is that the QGP has a small specific shear viscosity $\eta/s$, with its minimum near the QCD crossover inferred to be of order $0.1$~\cite{Bernhard:2019bmu,JETSCAPE:2020mzn} and thus compatible with the value $\eta/s=1/(4\pi)$ obtained for strongly coupled gauge theories~\cite{Kovtun:2004de}.

More recently, theoretical and experimental developments have pushed our understanding of hydrodynamics toward increasingly extreme conditions. Phenomenological descriptions of heavy-ion collisions require the system produced immediately after the collision to approach a regime governed by hydrodynamics within a remarkably short time, typically less than $1~\mathrm{fm}/c$~\cite{Kurkela:2015qoa}. Importantly, this rapid onset of hydrodynamic behavior, commonly referred to as hydrodynamization~\cite{Kurkela:2015qoa,Heller:2016rtz}, does not necessarily imply complete local thermalization or pressure isotropization: hydrodynamics may become applicable while the microscopic system remains highly anisotropic and far from equilibrium.

The puzzle has become even more striking with the observation of long-range correlations commonly associated with collective flow not only in heavy-ion collisions, but also in much smaller systems, including high-multiplicity proton--proton and proton--lead collisions~\cite{CMS:2010ifv,CMS:2012qk}, and, more recently, in light-ion collisions (oxygen--oxygen, neon--neon, neon--lead, etc.)~\cite{ALICE:2025luc,ATLAS:2025nnt,CMS:2025tga,LHCb:2025ixz}. 
Hydrodynamic calculations have achieved considerable success in describing these observations even in such small and short-lived systems~\cite{Weller:2017tsr}. A complementary perspective comes from ultracold atomic gases, where the interaction strength, particle number, and initial geometry are all under experimental control. There, anisotropic elliptic expansion characteristic of collective flow has even been observed in ensembles of as few as ten atoms~\cite{brandstetterEmergentInteractiondrivenElliptic2025}. These observations raise a fundamental question: are microscopic interactions sufficiently strong and frequent to drive collective hydrodynamic behavior on these remarkably short time and length scales~\cite{Kurkela:2018qeb}?

An important theoretical advance was the recognition that rapidly expanding relativistic systems can exhibit universal far-from-equilibrium dynamics~\cite{Heller:2015dha,Heller:2016rtz,Romatschke:2017vte}. When expressed in terms of appropriately scaled macroscopic observables, solutions initialized far from equilibrium and with widely different initial conditions rapidly approach common trajectories, known as hydrodynamic attractors, well before local thermal equilibrium is established~\cite{Enss:2026jps}. This early loss of sensitivity to the initial state provides a possible explanation for the rapid onset of hydrodynamic behavior and has made attractor dynamics a central framework for studying far-from-equilibrium evolution in heavy-ion collisions.

Most detailed studies of hydrodynamic attractors have focused on highly symmetric systems, particularly boost-invariant and transversely homogeneous Bjorken flow. Attractor behavior has also been investigated in inhomogeneous systems with transverse expansion using initial conditions based on optical Glauber models~\cite{Romatschke:2017acs,Behtash:2017wqg,Blaizot:2017lht,Strickland:2018ayk,Blaizot:2019scw}. However, these studies do not establish how generic transverse perturbations with a finite wave vector modify the approach to the attractor. 
Because transverse gradients drive the development of radial and anisotropic collective flow, understanding their influence is essential for assessing the phenomenological relevance of the attractor picture for small systems such as the quark--gluon plasma.

Kinetic theory provides a natural framework for addressing this question because it evolves the full one-particle distribution function and describes how hydrodynamic behavior emerges from microscopic phase-space dynamics. Since our primary interest is to isolate the effects of transverse gradients during the early far-from-equilibrium evolution, we employ the relaxation-time approximation (RTA) for the collision kernel~\cite{ANDERSON1974466,Baym:1984np,Florkowski:2013lya}. The RTA retains the competition between free streaming, expansion, and collisional relaxation while remaining sufficiently tractable to resolve the coupling among different angular moments induced by transverse gradients. It therefore allows us to isolate the consequences of broken transverse homogeneity without the additional complications of a realistic QCD collision kernel.

In this work, we take a first step toward systematically characterizing hydrodynamic attractors in kinetic theory at finite transverse wave number $k$. We generalize the adiabatic hydrodynamization framework of Refs.~\cite{brewer_adiabatic_2021,brewerFarfromequilibriumSlowModes2022, blaizot_fluid_2018} to include transverse spatial gradients and determine how the associated timescale $k^{-1}$ competes with longitudinal expansion and microscopic relaxation. We then investigate how transverse inhomogeneities modify the universal far-from-equilibrium evolution and establish the extent to which attractor behavior persists beyond the transversely homogeneous Bjorken-flow limit.

The paper is structured as follows: 
In Section~\ref{sec:kin-theory}, we outline a kinetic theory to model the system and express the evolution equation with general spatial transverse gradients in terms of an effective matrix equation with a spherical harmonic basis, while reviewing the adiabatic hydrodynamization framework, which allows for a simple analysis of attractors in terms of an eigenvalue problem. In Section~\ref{subsec:no-gradients-recap}, we review previous results using this framework neglecting transverse spatial gradients. We discuss the decoupling of the evolution equation into different sectors and differentiate between different attractors. In Section~\ref{sec:spatial-transverse-gradients}, we introduce the spatial transverse gradients and discuss their role in modifying the spectrum and the resulting attractors. We further connect our theory to the hydrodynamic limit at late times and investigate how the path toward thermalization is influenced by large gradients.

\section{Kinetic theory and setup}\label{sec:kin-theory}
\subsection{Evolution equation}
We treat the quark--gluon plasma as a collection of massless particles ($g_{\mu\nu} p^\mu p^\nu = 0$, $p = p^0$), described by a single-particle distribution function $f = f(\vec{r}, \vec{p}, t)$. For a boost-invariant system, the kinetic equation expressed in Bjorken coordinates is given by 
\begin{equation}
\left(\partial_\tau  - \frac{p_z}{\tau}  \partial_{p_z}  + \frac{ \vec{p}_\perp}{p}\cdot \nabla_\perp \right)f(\vec{x}_\perp, \vec{p}_\perp, p_z, \tau) = -\hat{C}[f],
\end{equation}
where $\tau = \sqrt{t^2 -z^2}$ is the boost-invariant proper time (see Appendix~\ref{appendix:derivation-boltzmann}), and $\hat{C}[f]$ is the collision term. 
We analyze the transverse gradients in Fourier space $\vec{k}_\perp = (k_x, k_y)$,
\begin{align}
f(\vec{k}_\perp, \vec{p}_\perp, p_z, \tau) \equiv \int \dd^2 \vec{x}_\perp e^{-i \vec{k}_\perp\cdot \vec{x}_\perp} f(\vec{x}_\perp, \vec{p}_\perp, p_z, \tau)
\end{align}
such that the equation in Fourier space is 
\begin{equation}
\left(\partial_\tau  - \frac{p_z}{\tau}  \partial_{p_z}  + i\frac{ \vec{k}_\perp\cdot\vec{p}_\perp}{p} \right)f(\vec{k}_\perp, \vec{p}_\perp, p_z, \tau) = - \hat{C}[f]\,.
\end{equation}
Expressing this equation in terms of the energy density distribution,
\begin{align}
F(\vec{k}_\perp, \cos\theta, \phi, \tau) \equiv  \frac{1}{(2\pi)^3} \int_0^{\infty} \dd{p} 4\pi p^2\times p f(\vec{k}_\perp, \vec{p}_\perp, p_z, \tau)\,,
\end{align}
with a change of variables into spherical coordinates in momentum space such that $\cos\theta = p_z/p$, $\tan\phi = p_y/p_x$, we write the equation in terms of a free-streaming part ($\hat{H}_\text{free}$), a gradient part ($\hat{H}_g(\vec{k})$), and a collision integral ($\hat{C}[F]$) as
\begin{align}\label{eq:F-evolution}
\begin{split}
\partial_\tau F
    &=  -\frac{1}{\tau} \hat{H}_{\text{free}} F- i \hat{H}_\text{g} F  - \hat{C}[F]\,,\\
    \hat{H}_{\text{free}}&\equiv \left(4 \cos^2 \theta - \sin^2 \theta \cos\theta\pdv{}{\cos\theta}\right)\,, \\
    \hat{H}_g&\equiv  \sin\theta (k_x \cos\phi + k_y \sin\phi) \,.
\end{split}
\end{align}

Following \cite{blaizot_fluid_2018, brewerFarfromequilibriumSlowModes2022}, we use a method of moments to solve equation~\eqref{eq:F-evolution}. We define the moments of the spherical harmonic expansion,
\begin{align}
L_{lm}(\vec{k}_\perp,\tau) \equiv \int \frac{\dd{\Omega}}{4\pi} (Y_l^m(\theta, \phi))^*  F(\vec{k}_\perp, \cos\theta, \phi, \tau)  \,,
\end{align}
where the measure is over the solid angle in momentum space and $Y_l^m$ are the spherical harmonic functions. We express our distribution $F$ as a (formally infinite) vector of moments, 
\begin{align}
\Psi = (L_{00}\,,\,L_{1,-1}\,,\,L_{1,0}\,,\,L_{1,1}\,,\,L_{2,-2}\,,\,L_{2,-1}\,,\,\dots)
\end{align}
where $l$ takes integer values from $0$ to infinity and $m$ takes integer values from $-l$ to $l$. This expansion is physically well motivated: through the definition of the stress-energy tensor in kinetic theory as $T^{\mu\nu} = \int \frac{\dd^3{\vectorbold{p}}}{(2\pi)^3 } \frac{p^\mu p^\nu}{p^0} f$, the first nine moments in this expansion with $l=0,1,2$ correspond directly to the components of the stress-energy tensor\footnote{Subject to the constraint $T_\mu^\mu = 0$ for conformal systems.}. In particular, the energy density can be computed as $T^{00} = \varepsilon = L_{00}$, while the momentum densities are given by linear combinations of the moments with $l=1$. Since energy and momentum are microscopically conserved quantities, we will call these first four moments with $l=0,1$ the \emph{hydrodynamic moments} and will sometimes denote them as 
\begin{align}
L_{(lm)_H} = \{L_{00}\,,\,L_{1,-1}\,,\,L_{1,0}\,,\,L_{1,1} \}\,.
\end{align}
The exact relations between the moments and the stress tensor are given in Appendix~\ref{appendix:stress-tensor-relations}. Expressing \eqref{eq:F-evolution} in the basis of spherical harmonics transforms it into a discrete set of coupled ordinary differential equations for the moments $L_{lm}$, which we may write as a Schr\"odinger-like\footnote{Note that the sign convention is not the same as in ordinary quantum mechanics.} equation $\partial_\tau \Psi = - \mathcal{H} \Psi$, where $\mathcal{H} = H_\text{free}/\tau + i H_g + C $ is the coupling matrix in the spherical harmonic basis. This physically motivated expansion and hierarchy in the moments justifies truncating this series of differential equations, allowing in practice for relatively uncomplicated numerical integration. Although the coupling matrix $\mathcal{H}$ is not Hermitian, we follow the now-standard abuse of language and continue to refer to it as the Hamiltonian.

We now turn to the collision term in \eqref{eq:F-evolution}. For simplicity, we use the relaxation-time-approximation (RTA)\footnote{Also known as the Bhatnagar--Gross--Krook (BGK) operator.} for the collision integral $\hat{C}[F] = (F-F_\text{eq})/\tau_R$, which forces the system to relax to local equilibrium over a timescale $\tau_R$. We further set $\tau_R = 1$ and express everything in corresponding dimensionless units\footnote{It was shown in \cite{blaizot2021attractor} that introducing power-law time dependence for the relaxation time of the form $\tau_R \sim \tau^{1-\Delta} $ does not alter the structure of the coupling equations after introducing a rescaled time parameter $w = \tau/\tau_R = (\tau/\tau_1)^{\Delta}$, where $\tau_1 = \tau_R(\tau_1)$.}. We consider elastic collisions that conserve energy and momentum, such that the hydrodynamic moments of $F_\text{eq}$ correspond exactly to those of $F$, or $\int \dd{\Omega} (Y_l^m)^* F = \int \dd{\Omega} (Y_l^m)^* F_\text{eq}$ for $l=0,1$. The simplest choice for $F_\text{eq}$ is therefore a distribution which only contains these hydrodynamic moments in the spherical harmonic expansion:
\begin{align}
    F_\text{eq} = \sum_{(lm)_H} L_{lm} Y_l^m = T^{00} + 3 (T^{0x} \hat{p}_x +T^{0y} \hat{p}_y+ T^{0z} \hat{p}_z ), \quad \hat{p}_i = p_i/p.
\end{align}
Putting everything together, the kinetic equation~\eqref{eq:F-evolution} for the moments $L_{lm}$ is explicitly given by
\begin{equation}\label{eq:coupling-matrix}
\boxed{\begin{aligned}
    \partial_\tau \Psi &= - \mathcal{H}\Psi, \quad \mathcal{H} = \frac{1}{\tau }H_\text{free} + i H_g + C, \\[10pt]
    H_\text{free} &=  \delta_{m,m'} \left( A_{l}^{m} \delta_{l',l} + B_{l}^{m} \delta_{l',l-2} + C_{l}^{m} \delta_{l',l+2} \right),\\[10pt]
   H_\text{g} &= \delta_{m',m-1} (k_x - i k_y) \left( D_{l'}^{-m'} \delta_{l',l+1}- D_l^m \delta_{l',l-1} \right)\\
    & + \delta_{m',m+1}(k_x + i k_y) \left(D_l^{-m} \delta_{l',l-1}-D_{l'}^{m'}\delta_{l',l+1} \right),\\[10pt] 
    C &= \delta_{m,m'}\delta_{l,l'} (1- \delta_{(lm),(lm)_H})/\tau_R
\end{aligned}}
\end{equation}
with numerical coefficients given by
\begin{equation}\label{eq:coupling-coefficients}
\begin{aligned}
A_l^m &=    \frac{7 l^2+7 l-5 m^2-4}{(2 l-1) (2 l+3)}, & \quad B_l^m &= \frac{l+2}{2 l-1} \sqrt{\frac{((l-1)^2-m^2) (l^2-m^2)}{(2 l-3) (2 l+1)}},\\
C_l^m &= B_{l+2}^m \left(\frac{5}{4+l}-1\right), & \quad D_l^m &= \frac{1}{2} \sqrt{\frac{(l+m-1) (l+m)}{(2l+1)(2l-1)}}.
\end{aligned}
\end{equation}
The derivation of these coefficients is given in Appendix~\ref{appendix:derivation-moment-eqns}. The coefficients $A,B,C$ for the free-streaming part $H_\text{free}$ and the collisions $C$ were already computed\footnote{These references use a slightly different normalization choice for the spherical harmonics, as explained in Appendix~\ref{appendix:derivation-moment-eqns}.} in Ref.~\cite{blaizot_fluid_2018} for the case $m=0$, and in Ref.~\cite{brewerFarfromequilibriumSlowModes2022} for generic $m$.

We conclude this section with a few general remarks on the structure of the system \eqref{eq:coupling-matrix}. Without loss of generality, we align our spatial coordinate system's $x$-axis with the direction of the gradient $k_y=0, k_x = k$. Note that in the absence of such gradients, the $SO(2)$ transverse rotational symmetry in position space of the Boltzmann equation leads to the decoupling of moments with different $m$, as can be seen from the $\delta_{m,m'}$ prefactor in $H_\text{free}$ and $C$. In addition, the $\mathbb{Z}_2$ symmetry $p_z \leftrightarrow -p_z$ of the Boltzmann equation further ensures that only moments of the same parity $s \equiv (-1)^{l+m}$ couple to each other, such that we have separate evolution equations $\partial_\tau \psi_{m,s} = -\mathcal{H}_{m,s} \psi_{m,s}$ for each sector $(m,s)$\footnote{In other words, $\mathcal{H}$ can be written in block-diagonal form $\mathcal{H} = \bigoplus_{m,s} \mathcal{H}_{m,s}$.}. We will use uppercase letters when referring to quantities of the full distribution (such as $\Psi$ for the full moment vector) and lowercase letters to denote quantities for a specific sector (such as $\psi$), dropping the $m$ subscripts for visual clarity and, unless otherwise specified, referring only to the even ($s=+$) sectors, as these are the main focus of this work. We call the sectors with $(m,s) = (0, \pm), (\pm1,+)$ \emph{hydrodynamic sectors}, given that they contain the hydrodynamic moments $L_{(lm)_H}$,
\begin{align}
\begin{split}
\psi_{0,+}(k,\tau) &= (L_{00}, L_{20}, L_{40}, \cdots)\,,\\
\psi_{0,-}(k,\tau) &= (L_{10}, L_{30}, L_{50}, \cdots)\,,\\
\psi_{1,+}(k,\tau) &= (L_{11}, L_{31}, L_{51}, \cdots)\,,\\
\psi_{-1,+}(k,\tau) &= (L_{1,-1}, L_{3,-1}, L_{5,-1}, \cdots)\,.
\end{split}
\end{align}
Note that, because the distribution function is real in position space, $F(\vec{x}_\perp, \cos\theta, \phi, \tau) \in \mathbb{R}$, the moments with negative $m$ are related by $L^*_{l,m}(k)= (-1)^m L_{l,-m}(-k)$, such that $\psi_{-1,+}(k,\tau)=-\psi_{1,+}^*(-k,\tau)$. All other sectors with $\abs{m}\geq 2$ are called \emph{non-hydrodynamic sectors}.
It is not difficult to check that the first four equations of \eqref{eq:coupling-matrix} in the hierarchy, which are the evolution equations for the hydrodynamic moments, correspond directly to the conservation of the stress-energy tensor $\grad_\mu T^{\mu\nu} = 0$.

The objective of this work is to investigate how the dynamics of the system change when this transverse gradient coupling term $H_g$ is included, and consequently how modes with different $k$ are driven toward thermalization.
We immediately note that transverse gradients break the original $SO(2)$ symmetry and therefore couple moments of neighboring $m$ sectors to each other. As we will show in Section~\ref{sec:population-sectors}, this coupling will significantly increase the importance of non-hydrodynamic moments in the expansion $F = \sum L_{lm} Y_l^m$ and thus modify attractors and the thermalization dynamics of the system.

\subsection{Adiabatic hydrodynamization}\label{subsec:adiabatic-hydro}
Before directly solving the Boltzmann equation, consider the general form of the equation $\partial_\tau \psi = - \mathcal{H} \psi$. Just as in ordinary quantum mechanics, we can infer important features of the dynamics by analyzing the eigenspectrum of the coupling matrix. 

If the coupling matrix is time-independent (static), one can find its complete basis of (right) eigenvectors $\mathcal{H}_{\rm static}\phi_n = \lambda_n \phi_n$. Performing a change of basis we can study the time evolution of the superposition coefficients in the eigenbasis,
\begin{align}
\psi(\tau) =& \sum_n a_n(\tau) \phi_n\,,\\
\partial_\tau a_n(\tau) =& - \lambda_n a_n(\tau)\,,
\end{align}
which is readily solved by
\begin{align}
a_n(\tau) =& \, a_n(\tau_0) \exp(-\lambda_n (\tau-\tau_0))\,,\nonumber\\
=& \, a_n(\tau_0) \exp(-\Re\{\lambda_n\} (\tau-\tau_0)-i\Im\{\lambda_n\} (\tau-\tau_0)).
\end{align}
We see immediately that the real parts of the eigenvalues $\lambda_n$ set the decay rate for their corresponding eigenmodes $\phi_n$, noting the different sign convention from quantum mechanics. If there is a significant gap in the (real) spectrum such that $\Re{\lambda_0} \ll \Re{\lambda_{n>0}}$, then all but the lowest eigenmode $\phi_0$, or ``ground state'',  will decay away quickly, such that $\psi(\tau) = \sum_n a_n(\tau) \phi_n \rightarrow a_0(\tau) \phi_0$ on a timescale $\Delta \tau\sim ( \min_{n>0}(\Re{\lambda_n})-\Re{\lambda_0})^{-1}$ and the system loses information about the initial conditions encoded in $a_{n>0}$. This reduction in the effective degrees of freedom of dynamical systems is commonly known as an attractor, and suggests that the system may be well described by an effective theory with just a few degrees of freedom, such as the moments contained in the ground state $\phi_0$.

In recent years, hydrodynamic attractors have received considerable attention as a possible explanation for the emergence of hydrodynamic behavior in the quark--gluon plasma while it remains far from equilibrium~\cite{Kurkela:2019set}. The essential complication for an expanding plasma is that the effective Hamiltonian is time-dependent. Nevertheless, at each proper time we can define its instantaneous right and left eigenvectors\footnote{Note that because $\mathcal{H}$ is not Hermitian, the left and right eigenvectors are not related to each other by simple Hermitian conjugation.} through
\begin{align}
    \mathcal{H}(\tau)\phi_n(\tau)
    &=\lambda_n(\tau)\phi_n(\tau)\,, \\
    \varphi_n^\dagger(\tau)\mathcal{H}(\tau)
    &=\lambda_n(\tau)\varphi_n^\dagger(\tau)\,,
\end{align}
with $\varphi_n^\dagger(\tau)\phi_m(\tau)=\delta_{nm}$. However, because the instantaneous eigenbasis itself evolves, the evolution of the expansion coefficients is not diagonal in this basis. Writing the state as $\psi(\tau)=\sum_n a_n(\tau)\phi_n(\tau)$, one obtains
\begin{align}
\partial_\tau a_n(\tau)=-\lambda_n(\tau) a_n(\tau)-\sum_m \mathcal{A}_{nm}(\tau)a_m(\tau),\qquad\mathcal{A}_{nm}\equiv\varphi_n^\dagger\partial_\tau\phi_m .
\end{align}
The off-diagonal components of $\mathcal{A}_{nm}$ describe transitions between instantaneous eigenmodes induced by the time dependence of the Hamiltonian. However, if the rate of change of the eigenbasis is small compared to the eigenvalue differences $\abs{\mathcal{A}_{nm}/ (\lambda_n-\lambda_m)} \ll 1$ for $n\neq m$, we can neglect the off-diagonal terms and still consider the system to be in its instantaneous ground state.
This framework is known as \emph{adiabatic hydrodynamization} and was introduced in Ref.~\cite{brewer_adiabatic_2021} and further developed in Refs.~\cite{rajagopalAdiabaticHydrodynamizationEmergence2025, deLescluze2025adiabatic,Rajagopal:2025nca}. In summary, the adiabatic hydrodynamization framework introduces an effective Hamiltonian to frame the emergence of attractors as the decay and subsequent evolution of a system to its instantaneous ground state. The real parts of the eigenvalues characterize the instantaneous decay rates, while a spectral gap separates the slowly evolving attractor mode from rapidly decaying excitations. 

We can define an attractor curve associated with a ratio of moments. For a given pair $(l',l)$ belonging to the same $m$ sector, consider
\begin{align}
    \frac{L_{l'm}(\tau)}{L_{lm}(\tau)}.
\end{align}
We say the ratio has an attractor curve if solutions from a wide range of initial conditions converge to a common function of $\tau$. Once the slowest instantaneous mode dominates, this universal ratio is determined by the components of its eigenvector,
\begin{align}
    \frac{L_{l'm}(\tau)}{L_{lm}(\tau)}
    \simeq
    \frac{
        [\phi_0^{(m)}(\tau)]_{l'}
    }{
        [\phi_0^{(m)}(\tau)]_{l}
    }.
\end{align}

\section{Review of dynamics without gradients}\label{subsec:no-gradients-recap}
We first review the dynamics for the $k=0$ case, which has previously been studied in Refs.~\cite{blaizot_fluid_2018,brewerFarfromequilibriumSlowModes2022, brewer_adiabatic_2021}, to facilitate understanding and comparison with the dynamics at finite transverse spatial gradients explored in Section~\ref{sec:spatial-transverse-gradients}. 
As noted in Section~\ref{sec:kin-theory}, at $k=0$ the system decouples into separate sectors labeled by $(m, s)$ where $s$ is the parity index indicating $l+m$ even or odd. We may therefore analyze the spectrum and dynamics of each sector separately.
At $k=0$, we may further only consider $m\geq 0$. In practice, even at finite gradients $k>0$, the $m$ and $-m$ sectors contain physically equivalent information (just as $k$ and $-k$ technically describe different gradients but correspond to physically equivalent situations). So while both positive and negative $m$ need to be included in solving equation~\eqref{eq:coupling-matrix}, we will often present results for $\abs{m}$ as opposed to positive and negative $m$ sectors separately.

\subsection{Solution for an individual $(m,s)$ sector at $k=0$}
\subsubsection{Non-hydrodynamic sectors ($\abs{m}\geq 2$)}\label{subsec:non-hydro-sectors}
We now move to solving the Boltzmann equation~\eqref{eq:coupling-matrix}. Note that for any non-hydrodynamic sector, the collision term is proportional to the identity matrix $I$, and the form of the coupling matrix in this case reduces to $\mathcal{H} \xrightarrow{k=0} H_\text{free}/\tau + I/\tau_R$, suppressing the $m$ indices for visual clarity.  The collision term $\propto I$ can then be eliminated using a simple rescaling $\psi \rightarrow \tilde{\psi} = e^{\tau/\tau_R} \psi$, $y \equiv \log{(\tau/\tau_R)}$, leading to
\begin{align}\label{eq:rescaling-non-hydro}
    \partial_\tau \psi  = - \left(\frac{1}{\tau}H_\text{free} + \frac{1}{\tau_R}I\right)\psi \iff \partial_y \tilde{\psi} = -H_{\text{free}} \tilde{\psi}.
\end{align}
The dynamics of the non-hydrodynamic sectors are therefore completely determined (up to some global rescaling) by the free-streaming terms of the Boltzmann equation. Concretely, if the eigenvectors of the time-independent free-streaming matrix satisfy $H_\text{free} \phi_n = \epsilon^{\text{free}}_n \phi_n$, then $\mathcal{H} \phi_n = (\epsilon^\text{free}_n/\tau +1/\tau_R)\phi_n$. That is, the eigenvectors of the free-streaming matrix are also the instantaneous eigenvectors of the full Hamiltonian. The eigenbasis of the Hamiltonian is therefore time-independent, and the adiabatic approximation discussed in Section~\ref{subsec:adiabatic-hydro} is exact. Analytically, the full solution of the evolution equation can be written
\begin{align}\label{eq:analytic-sol-non-hydro}
\psi(\tau) = \sum_n a_n(\tau){\phi}_n = e^{-(\tau-\tau_0)/\tau_R} \sum_n \left(\frac{\tau}{\tau_0}\right)^{-\epsilon_n^\text{free}} a_n(\tau_0) \phi_n.
\end{align}
The relaxation time $\tau_R$, which quantifies the collision rate, does not affect the dynamics but only enters as a global scale, just as expected from the rescaling in equation~\eqref{eq:rescaling-non-hydro}. Finally, we observe numerically that the spectrum of $H_\text{free}$ is bounded, $1/2<\Re(\epsilon^\text{free}_n) <7/4$, except for an isolated eigenvalue at $\epsilon^\text{free} = 2$ (we include an analytic proof of these bounds in Appendix~\ref{appendix:real-part-eval-bounds}). This sets bounds on the decay rate of all modes $L_{lm} \sim \tau^{-\Re{\epsilon^\text{free}_n}}$ at early times $\tau -\tau_0 \ll \tau_R$ before the exponential decay $e^{-(\tau-\tau_0)/\tau_R}$ of any non-hydrodynamic quantities becomes relevant.

\subsubsection{Hydrodynamic sectors}\label{subsec:hydro-sectors}

The collision matrix for the hydrodynamic sector, $C = \diag(0,1,1,\dots)/\tau_R$ differs from the identity by the projector onto the conserved moment. Because the collision matrix is no longer proportional to the identity, its effect cannot be removed by an overall $e^{\tau/\tau_R}$ rescaling. Instead, the relative importance of free streaming and collisions changes around
$\tau\sim\tau_R$, in sharp contrast to the non-hydrodynamic sectors discussed in the preceding section. However, computing the spectrum numerically we observe that it reamins very similar\footnote{This is unsurprising, as the difference between the hydrodynamic and non-hydrodynamic Hamiltonians is only a rank-one perturbation in the collision matrix.} to that of the non-hydrodynamic sectors, with $\epsilon_{n\geq1} \approx \epsilon_n^\text{free}/\tau + 1/\tau_R$. The only major difference is in the ground state eigenvalue, which, unlike the other modes, does not have a constant part and is approximately given by $\epsilon_0 \approx \epsilon_0^\text{free}/\tau$. 

This behavior can also be qualitatively deduced from the form of the collision matrix. At early times $\tau \ll \tau_R$, the dominant term in the Hamiltonian is the free-streaming term ($\propto 1/\tau$), so that the hydrodynamic sectors behave just like the previously discussed non-hydrodynamic sectors. At late times $\tau \gg \tau_R$, however, the collisions will dominate and the ground state vector will converge to $\phi_0 \propto (1,0,0\dots 0)$, so that it only contains the conserved hydrodynamic moment (e.g., $L_{00}$ in the $m=0$ sector).

The fact that there is exactly one conserved moment for each hydrodynamic sector leads to decaying attractor curves, in contrast to the constant attractor curves for the non-hydrodynamic sectors. As this will be relevant in Section~\ref{subsec:modification-attractors}, we emphasize this as a general feature: attractor curves for moment ratios $L_{l'm}/L_{lm}$ with $l'>l$ will always decay at late times for the hydrodynamic sectors $m=0,\pm1$ due to the single dominant mode at late times, while attractor curves for any moment ratios in the non-hydrodynamic sectors will be constant in time as a consequence of the time-independent ground state.

\begin{figure}[p]
    \begin{subfigure}{0.5\linewidth}
        \centering
        \includegraphics[width=\linewidth]{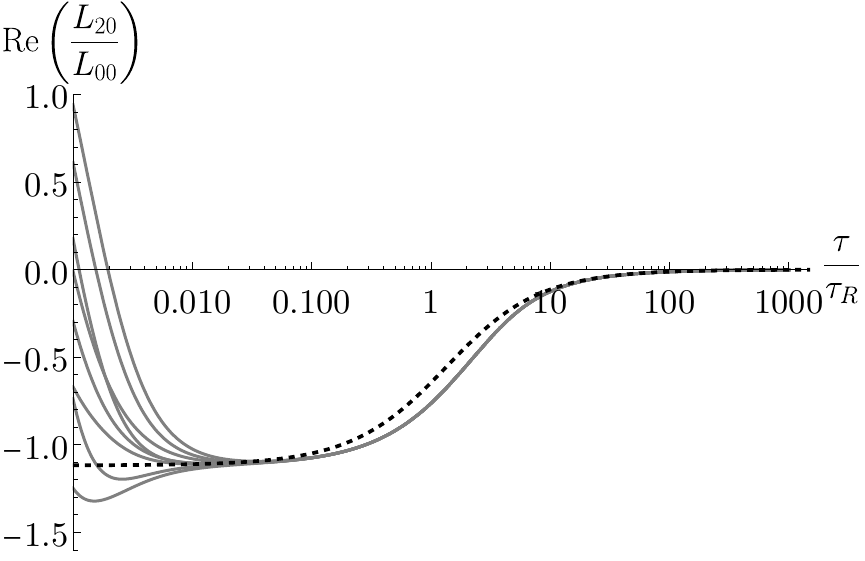}
        
    \end{subfigure}
    \begin{subfigure}{0.5\linewidth}
        \centering
        \includegraphics[width=\linewidth]{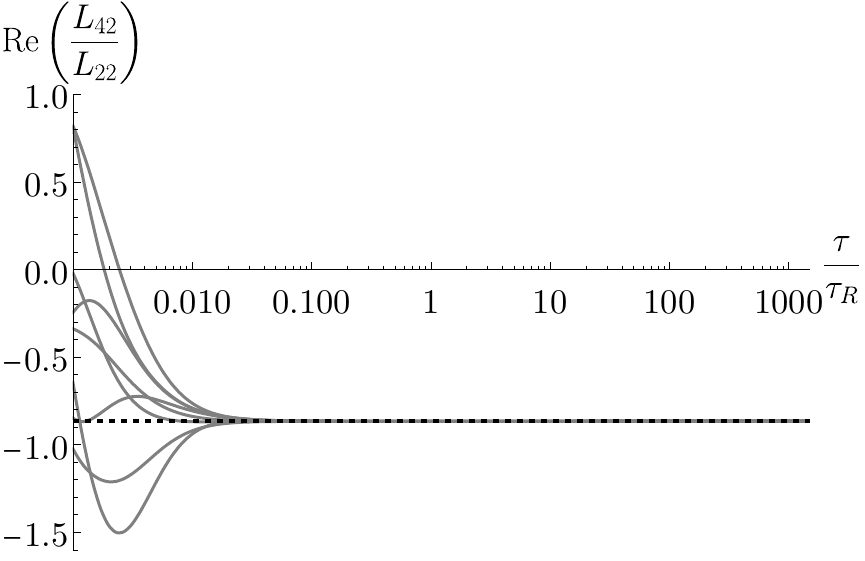}
    \end{subfigure}
    
    \caption{Attractor in the hydrodynamic sector $m=0$ (left) and in the non-hydrodynamic sector $m=2$ (right), both in the even ($s=+$) sector. Due to a significant gap in the spectrum, many initial conditions quickly converge to a universal solution, corresponding to the instantaneous ground state of the system, overlaid as a black dashed curve. The non-hydrodynamic sector has a time-independent eigenbasis, and as a consequence the ground state attractor (black dashed, right plot) is constant. In contrast, the ground state attractor curve for the hydrodynamic sector (black dashed, left plot) changes in time and decays. Due to the time-dependent basis in the hydrodynamic sector, the adiabatic approximation is not exact and higher eigenmodes are excited around $\tau \sim \tau_R = 1$, accounting for the discrepancy between the dashed and full lines (also see Figure~\ref{fig:k=0-eigendecomp}). This computation retains the first 251 moments of the sector hierarchy; for the $m=0$ sector this corresponds to moments up to $L_{500,0}$. The same results have previously been shown in Ref.~\cite{brewerFarfromequilibriumSlowModes2022}.}
    \label{fig:attractors-k0}
\end{figure}

\begin{figure}[p]
    \centering
    \includegraphics[width=\linewidth]{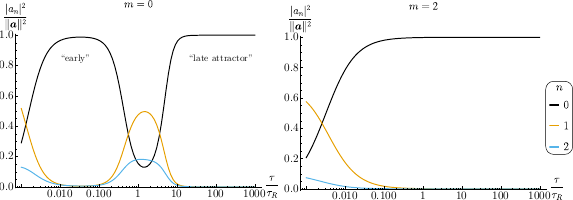}
\caption{\label{fig:k=0-eigendecomp}Overlap of the state with its instantaneous eigenstates. The first few coefficients of the decomposition in the eigenbasis $\psi = \sum a_n \phi_n$ are shown for the $m=0$ (left) and $m=2$ (right) sectors. Coefficients are normalized such that $\sum_n \abs{a_n}^2/\norm{a}^2 =1$. We clearly see how at early times $\tau \sim 0.03 \tau_R$, the system has already relaxed to the eigenmode $\phi_0$, at which point all sensitivity to the initial conditions is lost. Around the relaxation time, the adiabatic approximation fails, leading to transient excitations of higher eigenstates. For the non-hydrodynamic sector $m=2$, the eigenbasis is constant in time such that the adiabatic approximation is exact and no transient excitations are present. As further indicated in the left panel, in the hydrodynamic sector, we can differentiate between the early (non-hydrodynamic) attractor and the late (thermalization) attractor that only occurs after the relaxation time $\tau_R$. In contrast, the $m=2$ sector only contains an early, non-hydrodynamic attractor.}
\end{figure}

To illustrate these results and summarize the discussion in this section and Section~\ref{subsec:non-hydro-sectors}, we show in Figure~\ref{fig:attractors-k0} the ratios of the first two moments, $L_{20}/L_{00}$ and $L_{42}/L_{22}$, in the $m=0$ and $m=2$ sectors, respectively. The calculation closely follows Ref.~\cite{brewerFarfromequilibriumSlowModes2022}. In both sectors, solutions with a wide range of initial conditions rapidly converge onto a common curve much earlier than the relaxation time, $\tau \ll \tau_R$, demonstrating attractor behavior. The dashed curves show the adiabatic attractor curve predictions obtained from the corresponding ratios of components of the instantaneous ground-state eigenvector. In the $m=2$ sector, the eigenbasis is time-independent, and the resulting attractor ratio is constant and agrees well with the full numerical evolution. In the $m=0$ sector, the universal curve instead decreases and approaches zero as $\tau\to\infty$. A finite discrepancy arises at $\tau \sim \tau_R$ between this converged numerical curve and the instantaneous ground-state prediction. The origin of this discrepancy, which is associated with the time dependence of the eigenbasis and the resulting excitation of higher instantaneous modes, is examined in Section~\ref{subsubsec:eigendecomp-k0}.

The moment ratio shown for the $m=0$ sector also has a direct physical interpretation: it measures the quadrupole anisotropy of the momentum distribution and determines the longitudinal pressure through
\begin{align}
\frac{p_L}{\varepsilon}=\frac{1}{3}\left(1+\frac{2}{\sqrt{5}}\frac{L_{20}}{L_{00}}\right).
\end{align}
Consequently, the late-time decay $L_{20}/L_{00}\longrightarrow0$ corresponds to the isotropization of the longitudinal pressure, $p_L/\varepsilon\longrightarrow 1/3$,
as expected for an isotropic conformal system.

\subsubsection{Eigendecomposition}\label{subsubsec:eigendecomp-k0}
We further verify the applicability of the adiabatic hydrodynamization framework by directly computing the overlap (inner product) of the state $\psi(\tau)$ with each of the instantaneous eigenstates of the sector, showing how the ground-state fraction grows with time. For this, we recall the setup from Section~\ref{subsec:adiabatic-hydro}, and decompose $\psi(\tau) = \sum_n a_n(\tau) \phi_n(\tau)$, where the $\phi_n$ are the instantaneous eigenvectors of the sector Hamiltonian $\mathcal{H}$, consistently ordered according to their decay rates $\Re{\epsilon_n}$. Defining the coefficient vector $\boldsymbol{a}$ with elements $a_n$, and the matrix $P$ containing all $\phi_n$ as its columns $P = (\phi_0, \phi_1, \dots, \phi_N)$, we compute the decomposition $\boldsymbol{a} = P^{-1} \psi$. 
We then define the (fractional) \emph{occupation} of the $n$th eigenmode $\abs{a_n}^2/\norm{\boldsymbol{a}}^2$ as the squared overlap coefficient of that mode with the state, normalized such that they sum to one\footnote{In quantum mechanics language, this would be $\abs{\braket{\varphi_n}{ \psi}}^2/\braket{\psi}{\psi}$.}. 

In Figure~\ref{fig:k=0-eigendecomp}, we show the first three $\abs{a_n}^2/\norm{\boldsymbol{a}}^2$ of the $(m,s)=(0,+)$ and $(m,s)=(2,+)$ sectors as a function of the rescaled time $\tau/\tau_R$.
For the non-hydrodynamic $m=2$ sector, we see how the presence of the gap leads all higher modes to decay after a short time, such that the system is effectively in its ground state and $\abs{a_0}^2/\norm{\boldsymbol{a}}^2  \rightarrow 1 \implies \psi \propto \phi_0$. 

The hydrodynamic $m=0$ sector exhibits two distinct intervals of attractor behavior, one at early times and the other at late times. In both regimes, the instantaneous mode decomposition of $\psi$ is almost entirely dominated by the ground state. The two regimes are separated by a crossover around $\tau\sim\tau_R=1$, where the competition between free streaming and collisions causes the instantaneous eigenbasis to rotate rapidly. The rate of this basis rotation becomes comparable to the spectral separation between the ground state and the excited modes, so that the adiabaticity condition
$\left|\varphi_n^\dagger\partial_\tau\phi_0\right|\ll\left|\epsilon_n-\epsilon_0\right|$
is temporarily violated. These excitations are transient: once the eigenbasis begins to vary slowly again, the excited modes decay and ground-state dominance is restored, giving rise to the second attractor regime\footnote{We note here that the exact form of Figure~\ref{fig:k=0-eigendecomp} has a significant truncation dependence and should therefore only be interpreted qualitatively, which can be explained as a feature of high-dimensional vector decompositions in non-orthogonal bases, combined with the fact that the spherical harmonic basis is not in fact well suited to the free-streaming fixed point at early times.}. This further accounts for the discrepancy between the dashed ground state curve and the actual attractor curve shown in the left of Figure~\ref{fig:attractors-k0}, which was noted in the previous section.

\subsection{Comparison of sectors}\label{subsec:sector-pop-k0}
So far we have focused on the dynamics within a single sector. We now consider the full distribution function and compare the sectors to each other. Since the sectors are decoupled and evolve independently, a gap between states of different sectors does not directly affect the evolution of either sector. However, such a gap is still highly relevant for the complete distribution function $F$, as it will determine which sectors and moments contribute most strongly to the distribution function at any given time.
\subsubsection{Sector decomposition}
To quantify the sectors' contribution to the state, we decompose the moment vector into mutually orthogonal sector vectors $\Psi = \sum_m \psi_m$, where each $\psi_m$ contains nonzero entries only for moments of the fixed $m$. For example, $\psi_1  = (0, 0, L_{1,1}, 0,\dots,0, L_{3,1}, 0, \dots)$. We then define the population fraction of each sector by $r_m \equiv \norm{\psi_m}^2/\norm{\Psi}^2$, such that by construction $\sum_m r_m=1$. In particular, as already discussed in Sections~\ref{subsec:non-hydro-sectors} and~\ref{subsec:hydro-sectors}, the hydrodynamic sectors have ground state eigenvalues $\epsilon \propto  1/\tau$ compared to those of the non-hydrodynamic sectors, $\epsilon \simeq (\text{const.}/\tau +1/\tau_R)$. Therefore, around the relaxation time $\tau \sim \tau_R$, an increasingly large gap opens between the power-law decaying ground states of the hydrodynamic sectors ($\epsilon \propto 1/\tau$) and the exponentially decaying non-hydrodynamic states ($\epsilon \propto 1/\tau_R$). The contribution of the hydrodynamic sectors in the distribution function should therefore start to sharply increase after the relaxation time, corresponding to the hydrodynamization of the system.

\begin{figure}
    \begin{subfigure}{0.5\linewidth}
        \centering
        \includegraphics[width=\linewidth]{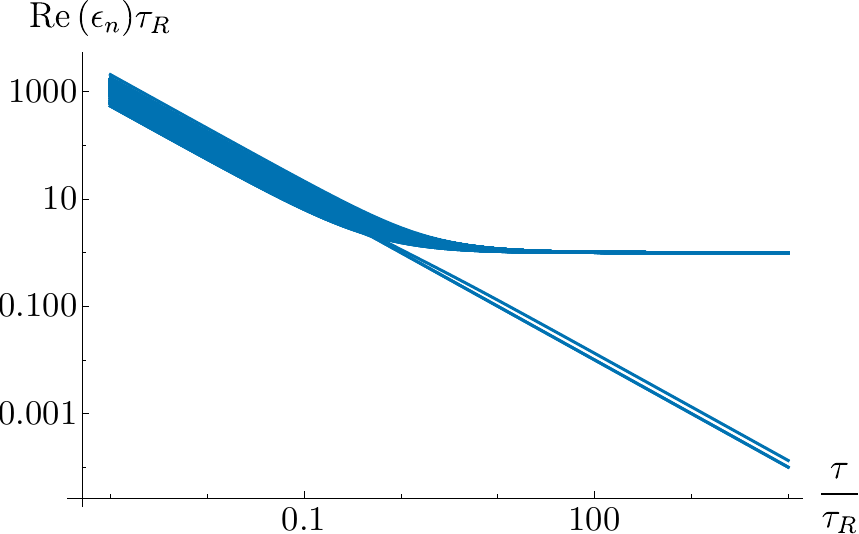}
        \caption{}
        \label{fig:Re-vs-time-full-k=0}
    \end{subfigure}
    \begin{subfigure}{0.5\linewidth}
            \centering
            \includegraphics[width=\linewidth]{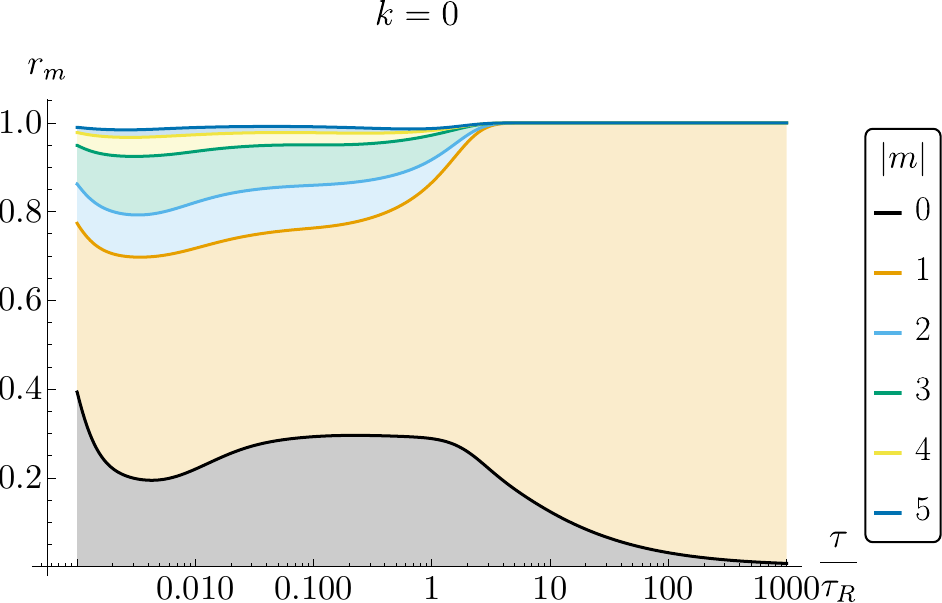}
            \caption{}
            \label{fig:sector-population-k0}
    \end{subfigure}
    \caption{\textbf{(a)}: Full (real) eigenspectrum of $\mathcal{H}$ at $k=0$. After $\tau \sim \tau_R$, an increasingly large gap opens between the power-law decaying ground states of the hydrodynamic sectors ($\epsilon \propto 1/\tau$) and the exponentially decaying non-hydrodynamic states ($\epsilon \propto 1/\tau_R$). The gaps at early times are not visible when considering many decoupled sectors together. \textbf{(b)}: Stacked plot of the sector population fraction $r_m \equiv \norm{\psi_m}^2/\norm{\Psi}^2$ for non-interacting/decoupled sectors at zero transverse gradients. While the sector contributions remain of the same order as in the initial state during the early (free-streaming) evolution of the system, shortly after the relaxation time $\tau >\tau_R$, the increasingly large gap between the hydrodynamic and non-hydrodynamic sectors leads to exponentially decaying non-hydrodynamic contributions, such that only the $m=0,\pm 1$ sectors remain. The small gap between the $m=0$ and $m=\pm 1$ sectors leads to a dominance of the $m=\pm1$ ground state at $\tau\gg 1$.}
\end{figure}
\subsubsection{Numerical analysis}
\paragraph{Spectrum}
We show the full time-dependent spectrum of the Hamiltonian $\mathcal{H}$ in Figure~\ref{fig:Re-vs-time-full-k=0}. We see many curves forming a band at early times, each decaying with the expected $1/\tau$ behavior. The gaps within each sector leading to the attractors discussed previously are not visible when showing all the sectors together. After the relaxation time, we observe two curves continuing to decay, while all other curves flatten out at $\epsilon = 1/\tau_R =1$, creating an increasing gap between the hydrodynamic and non-hydrodynamic sectors. The decaying curves correspond to the three eigenvalues of the ground states in each of the hydrodynamic sectors $m=0, \pm 1$, where the two $m=\pm 1$ eigenvalues are degenerate in their real part and therefore only visible as a single curve. The flat curves correspond to the eigenvalues of all other states, which decay exponentially after the relaxation time. We can also see a small but clearly resolved gap between the two hydrodynamic eigenvalue curves, where the lower one corresponds to the $m=\pm1$ ground state eigenvalue $\epsilon_0^{\abs{m}=1, \text{free}} = 1$, and the upper one to the $m=0$ ground state eigenvalue $\epsilon_0^{m=0, \text{free}} = \frac{4}{3}$.
\paragraph{Sectors}
In Figure~\ref{fig:sector-population-k0}, we show the population fraction of the different sectors computed with a randomly chosen initial condition, shown as a stacked plot to emphasize that the populations sum to 1. At early times we see that the population of the sectors remains roughly of the same order as in their initial state. This is a consequence of the absence of any significant gaps between the different sectors at early times, as seen from the continuous band of eigenvalues in Figure~\ref{fig:Re-vs-time-full-k=0}. At late times, we can clearly see the effect of the gap: all sector populations $r_{\abs{m}\geq 2}$ decay exponentially fast, and only the three hydrodynamic sectors $m=0,\pm1$ remain. Even later, when $\tau \gg 1$, the small gap between the $m = \pm 1$ and the $m=0$ sectors leads to a power-law decay of the $m=0$ sector. This is consistent with well-known results for relativistic Bjorken flow: as discussed in Section~\ref{subsec:hydro-sectors}, the ground states of the hydrodynamic sectors are states only containing the first (hydrodynamic) moment of that sector, that is, $L_{00}$ and $L_{1,\pm 1}$, corresponding to the energy density $T^{00}$ and the momentum densities $T^{0x}, T^{0y}$, respectively (see Appendix~\ref{appendix:stress-tensor-relations}). In Bjorken flow, these have time dependence $T^{00} \propto \tau^{-4/3}$ and $T^{0x} \propto \tau^{-1}$, such that at late times the momentum densities dominate the distribution function $F$ and the relative contribution of the energy density decays as $\tau^{-1/3}$.

\subsection{Size of the gap}\label{subsec:significant-gap}
In Section~\ref{subsec:adiabatic-hydro}, we discussed that attractors will occur if a significant gap is present in the real eigenspectrum of the Hamiltonian. We conclude this section by quantitatively specifying what we mean by ``significant''. Consider a state containing only the ground state and the first excited mode of some arbitrary sector, $\psi(\tau) \sim a_0(\tau) \phi_0 + a_1(\tau) \phi_1$. Defining the gap $\Delta \equiv \Re(\epsilon_1 - \epsilon_0)$, and considering the formal solution of the evolution equation of the non-hydrodynamic sectors \eqref{eq:analytic-sol-non-hydro}, we have
\begin{align}
    \frac{a_1(\tau)}{a_0(\tau)} = \frac{a_1(\tau_0)}{a_0(\tau_0)} \left(\frac{\tau}{\tau_0}\right)^{-\Delta}.
\end{align}
We arbitrarily define the attractor to be approached on a timescale when $\left(\frac{\tau}{\tau_0}\right)^{-\Delta} = \frac{1}{2}$, that is, when the overlap ratio between the ground and excited states has halved. In our simulations, we usually use $\tau_0 = 0.001 \tau_R$, and we therefore take a significant gap to mean $\Delta> \log(2)/\log(1000) \sim 0.1$. For a gap smaller than this, the attractor can only be reached far later than $\tau \sim 1000\tau_0 =\tau_R$, such that any early-time attractor features are not visible. 

\begin{figure}
    \begin{subfigure}{0.45\linewidth}
        \centering
    \includegraphics[width=\linewidth]{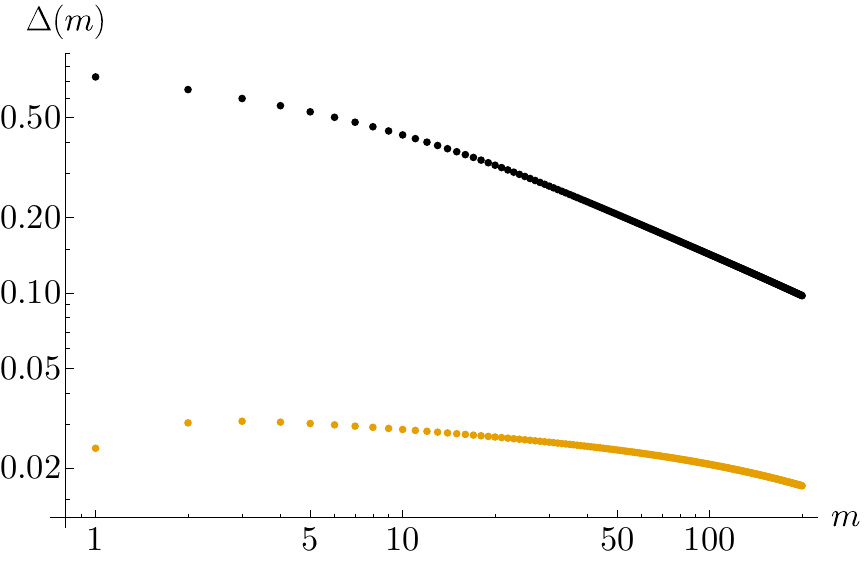}
    \caption{}
    \label{fig:delta-vs-m}
    \end{subfigure}
    \begin{subfigure}{0.55\linewidth}
        \centering
    \includegraphics[width=\linewidth]{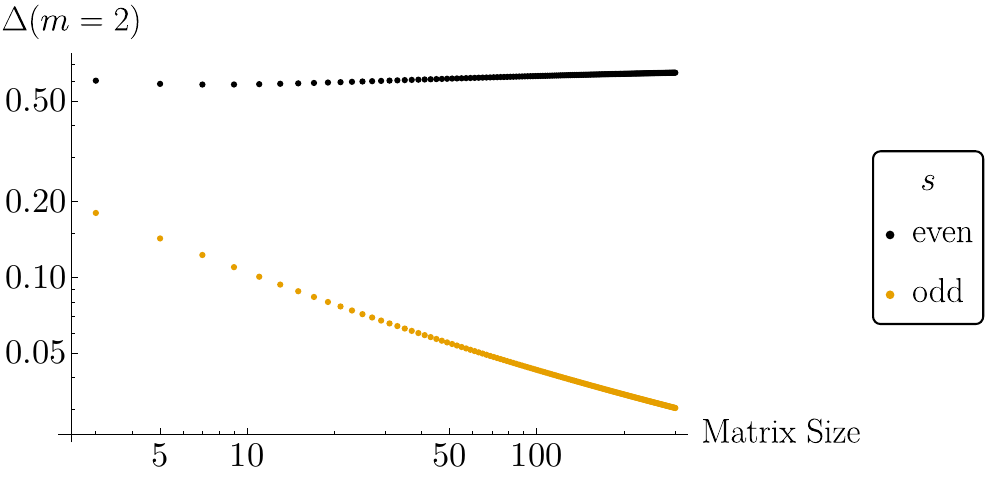}
    \caption{}
    \label{fig:delta-vs-trunc}
    \end{subfigure}
\caption{\label{fig:gap-quantitative} \textbf{(a)}: Numerical value of the gap $\Delta$ for each sector $m$, in black for the even ($s=+$) and orange for the odd ($s=-$) sectors. The even sectors all contain a significant gap $\Delta \gtrsim0.1$, whereas all odd sectors do not have a significant gap. Computed with truncated matrix size $N=301$. \textbf{(b)}: The gap of an example ($m=2$) sector (even and odd), as a function of the truncation size $N$. We see that the odd-sector gap is further reduced as the truncation is improved. In Appendix~\ref{appendix:trunc-dependence-gap}, we show that the odd-sector gap closes, $\Delta \to 0$, logarithmically slowly in the limit $N \rightarrow \infty$, and the even-sector gap approaches $3/4$ for any $m$ sector.}
\end{figure}

We show the numerically computed gap $\Delta$ for different sectors as a function of the sector $m$ in Figure~\ref{fig:gap-quantitative}. We see that all even sectors contain a significant gap between the ground state and first excited state, whereas all the odd sectors have very small gaps. The exact size of the gap decreases as the sector index $m$ is increased, such that higher $m$ sectors only reach their respective ground states at later times, although this effect only becomes noticeable for physically irrelevant $m \gtrsim 100$. At the same time, we observe that improving the numerical truncation leads to a slight increase of the gap in the even sectors, in contrast to a further decrease in the gap for the odd sectors. In Appendix~\ref{appendix:trunc-dependence-gap}, we include an analytical argument that shows how the odd sectors are actually gapless ($\Delta^{(-)} = 0$), and that the finite gap only arises through a finite truncation error in the basis. In contrast, all even sectors contain genuine gaps. Attractors are thus generically present in all even sectors but not in the odd sectors\footnote{The odd sector $(0,-)$ is the only exception to this and will display a late-time attractor after $\tau > \tau_R$. This is because it is also a hydrodynamic sector which develops a large gap between the ground state corresponding to the conserved moment $L_{10}$ and all other modes of the sector. The dynamics at late times are equivalent to those of the even hydrodynamic sectors, which are discussed in Section~\ref{subsec:hydro-sectors}.}. Concretely, moment ratios such as $L_{42}/L_{22}$ display attractor behavior, but not $L_{52}/L_{32}$. The even sector also holds most of the physically relevant information, since we expect the particle distribution function resulting from a heavy-ion collision to be roughly symmetric under the exchange $p_z \leftrightarrow -p_z$. Unless otherwise specified, we will therefore focus on the even sector of the distribution function ($s = +$) for the remainder of this work.

\section{Spatial transverse gradients}\label{sec:spatial-transverse-gradients}
We now turn to the influence of spatial transverse gradients on the spectrum, attractor features, and dynamics of the massless boost-invariant plasma. Without loss of generality, we align our coordinate axes such that $k_y = 0, k_x = k$. From inspecting equation~\eqref{eq:coupling-matrix}, we can see that at early times $\tau\ll 1/k, \tau \ll \tau_R$, the free-streaming terms $H_\text{free}/\tau$ will dominate the equation and the results of the previous section remain unchanged; it is only when $\tau$ is roughly of the order of the timescale $\tau_g\equiv k^{-1}$ that the gradient term becomes comparable to the other timescales in $\mathcal{H}$ and can thus change the dynamics of the system. Therefore, the system is still well described by (almost) decoupled sectors labeled by $(m,s)$ at early times, just as before, with these sectors gradually starting to mix around this gradient timescale $\tau_g$.  In Section~\ref{subsec:modification-attractors}, we will make this notion more precise and show that the actual timescale on which the dynamics are modified within any given sector depends only logarithmically on the gradient.

This section is structured as follows: We begin by analyzing the spectrum in the late-time (hydrodynamic) limit $\tau \gg \tau_R$ and find the new, global ground state modes of the system. We then discuss the relative contribution of hydrodynamic moments to the distribution function at late times. Mirroring Section~\ref{subsubsec:eigendecomp-k0}, we decompose the moments into the eigenstates of the system, and show the applicability of the adiabatic hydrodynamization framework, explicitly showing the transition from the decoupled sector ground states to the global ground states. Finally, we show how this interpolation between early and late ground states modifies the attractors in certain moment ratios.

\subsection{The hydrodynamic limit}\label{subsec:eigenspectrum-gradients-hydro-limit}

\begin{figure}
    \centering
    \includegraphics[width=0.55\linewidth]{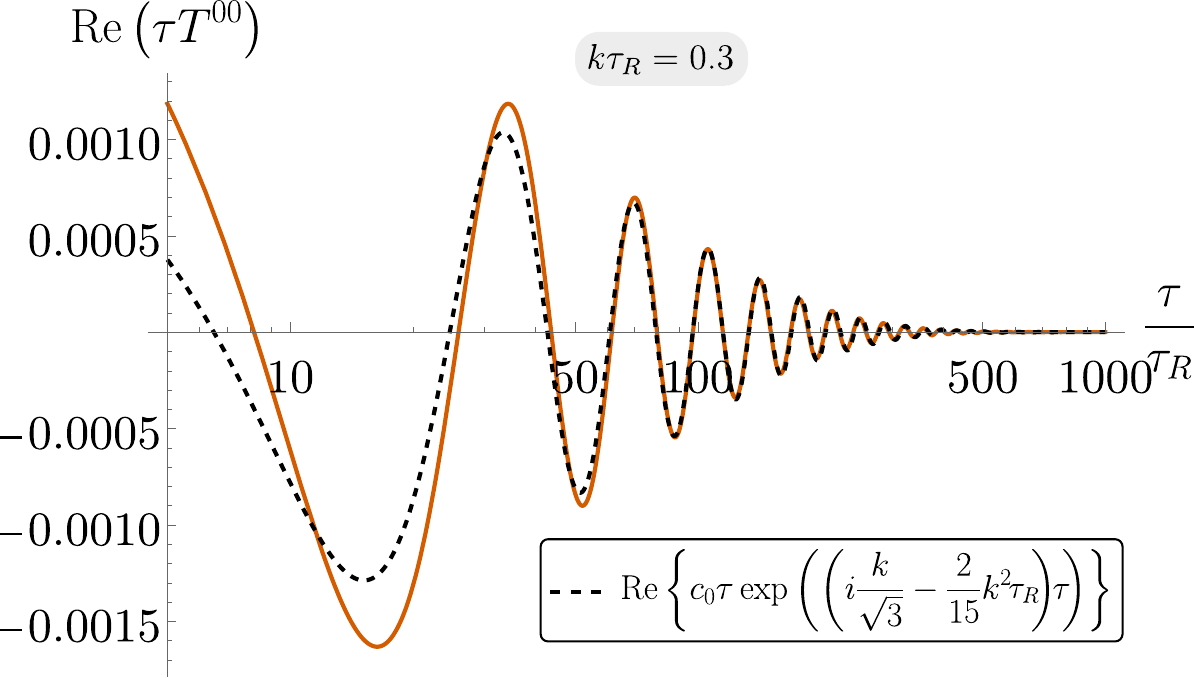}
    \caption{Late-time behavior of the energy density at a finite gradient $k \tau_R =0.3$. We can clearly see that the solution approaches the sound mode $T^{00} \propto \exp(- \epsilon_0\tau) $, with $\epsilon_0 = i k/\sqrt{3} + 2 k^2\tau_R/15$ as in equation~\eqref{eq:sound-dispersion}. Here we only fitted the complex constant $c_0 = (1.4 +0.67 i)\cdot 10^{-3}$ to match the solution. Observing this sound mode solution explicitly at late times is a strong indication that the system is indeed in the global ground state $\Phi_{0}$, as expected in the adiabatic hydrodynamization framework.}
    \label{fig:T00-vs-time}
\end{figure}

\paragraph{Small gradients}
In the late-time limit $\tau \rightarrow \infty$, we can directly compute the eigenvalues of the coupling matrix as a series expansion in $k$. In particular, we are interested in the dominant slow/hydrodynamic modes, or ground states, for which $\epsilon(k=0)= 0$ at $\tau \rightarrow \infty$. We find
\begin{align}
       \epsilon_{0,1} 
       \tau_R &= \pm i \frac{ k\tau_R}{\sqrt{3}}+\frac{2 (k\tau_R)^2 }{15} \pm i \frac{8 (k\tau_R)^3}{75 \sqrt{3}} - \frac{68 (k \tau_R)^4}{1575} + \order{(k\tau_R)^5}\label{eq:sound-dispersion} \\
       \epsilon_{2}\tau_R &= \frac{(k\tau_R)^2}{5} - \frac{(k\tau_R)^4}{175} + \order{(k\tau_R)^6}\label{eq:shear-dispersion}
\end{align}
and identify these as sound and shear modes, respectively. Specifically, the sound dispersion relation for a relativistic fluid in the hydrodynamic gradient expansion is given by  the BRSSS result~\cite{baier_relativistic_2008} $\omega =-i\epsilon = \pm c_s k - i \Gamma k^2 \pm \Gamma/c_s \left(c_s^2 \tau_R - \Gamma/2\right)k^3 + \order{k^4}$, with $\Gamma \equiv \frac{d-2}{d-1} \frac{\eta}{\varepsilon+P}$ and $c_s = 1/\sqrt{d-1}$ as the sound speed in a conformal fluid in $d$ spacetime dimensions. The shear viscosity for Bjorken flow in RTA was calculated in \cite{chattopadhyay_higher_2018} as $\eta/\tau_R = 4P/5$, such that for $d=4$ and inserting the equation of state for an ultrarelativistic fluid $\varepsilon = (d-1)P$, we get $\Gamma = 2/15$ and $\Gamma/c_s \left(c_s^2\tau_R - \Gamma/2\right) = 8/(75 \sqrt{3})$, exactly matching the coefficients of our series expansion. This demonstrates that our kinetic theory precisely matches the traditional hydrodynamic gradient expansion in the limit $\tau \rightarrow \infty$. 

We further find that in the limit $k\rightarrow 0$, these ground state eigenvectors are $\Phi_{0,1} = (\sqrt{2},\mp 1,\pm1,0,0,\dots,0)$ in the moment basis $\{L_{lm}\}$, which corresponds to the linear combination $T^{00} \pm \sqrt{3} T^{0x}$ of the stress-energy tensor components (see Appendix~\ref{appendix:stress-tensor-relations}), explicitly confirming this as a sound mode mixing the energy and momentum components of the stress-tensor. The shear mode is given by the eigenvector $\Phi_2 = (0,1,1,0,\dots,0)$ corresponding to $T^{0y}$.

Numerically solving the evolution equation~\eqref{eq:coupling-matrix}, we can further verify that the system at late times is indeed dominated by a sound mode. In Figure~\ref{fig:T00-vs-time}, we show the numerically computed behavior of the Fourier-transformed energy density at late times for a small value of $k$, clearly showing the sound mode oscillatory solution at frequency $k/\sqrt{3}$ and exponential decay constant $2k^2/15$. The apparent convergence of the energy density to the sound mode solution in Figure~\ref{fig:T00-vs-time} strongly suggests that the full state has indeed decayed to these global ground states $\Phi_{0,1,2}$, as the adiabatic hydrodynamization framework would predict. We will explicitly confirm this in Section~\ref{subsec:sector-gs-to-global-gs}.

\begin{figure}
\begin{subfigure}{0.5\textwidth}
    \centering
    \includegraphics[width=\linewidth]{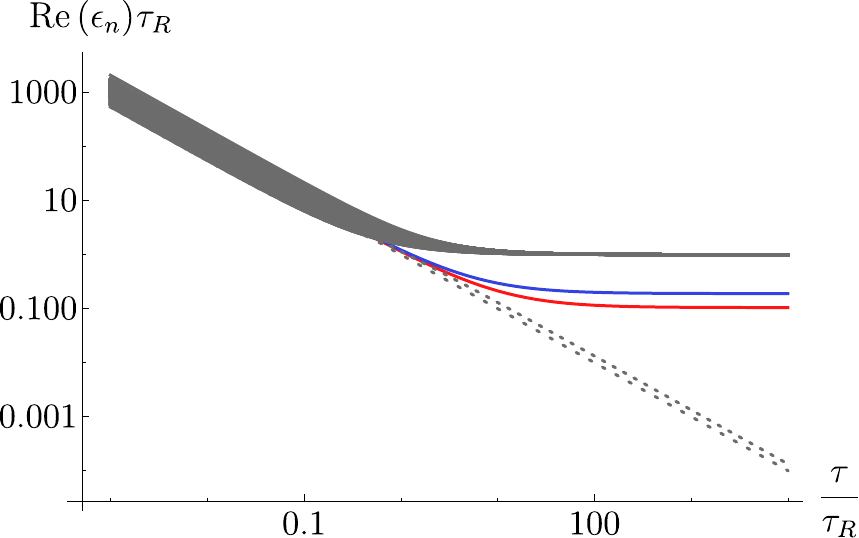}
    \caption{}
    \label{fig:Re-late-vs-time} 
\end{subfigure}
\begin{subfigure}{0.5\textwidth}
    \centering
    \includegraphics[width=\linewidth]{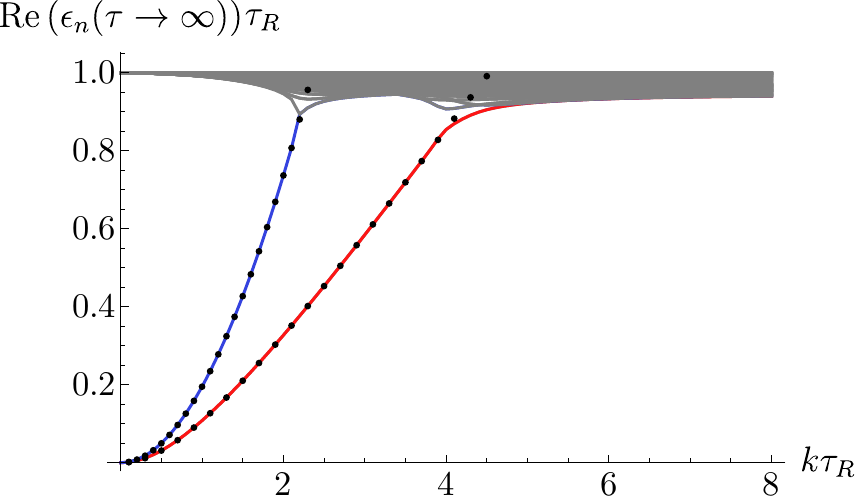}
    \caption{}
    \label{fig:Re-late-vs-k} 
\end{subfigure}

\caption{\label{fig:evals-k-dependence}
    \textbf{(a):} Time-dependent eigenspectrum at $k \tau_R = 1$, with the non-hydrodynamic modes shown in gray, and the hydrodynamic sound and shear modes shown in red and blue, respectively. The hydrodynamic modes at zero gradients are shown as a dashed curve. We see that for the most part, the spectrum at finite gradients is unchanged (compare to Figure~\ref{fig:Re-vs-time-full-k=0}), except that the hydrodynamic eigenvalues acquire a constant real part at late times. \textbf{(b):} Eigenspectrum at $\tau \rightarrow \infty$ as a function of $k \tau_R$. The black dots are computed from the poles of the exact sound and shear retarded correlator functions in Ref.~\cite{Romatschke:2015gic}, demonstrating that our framework precisely matches the traditional hydrodynamic gradient expansion. The finite truncation of the Hamiltonian leads to the spread around $\epsilon_n= 1/\tau_R$. At small $k \tau_R$ one can clearly see a $k^2$ dependence predicted from the series expansion in equation~\eqref{eq:sound-dispersion}. At large gradients $k \tau_R\gtrsim 4.53$, the gap between the hydrodynamic and non-hydrodynamic modes closes and the system no longer thermalizes via the hydrodynamic attractor path.}
\end{figure}

\paragraph{Large gradients}
For generic times and gradients, we can compute the eigenvalues numerically, shown in Figure~\ref{fig:evals-k-dependence}, and check that the early- and late-time limits are consistent with our analysis. Figure~\ref{fig:Re-late-vs-time} shows the spectrum as a function of time for some fixed value of $k$. We see that the spectrum looks almost identical to the $k=0$ spectrum shown in Figure~\ref{fig:Re-vs-time-full-k=0}. Only the hydrodynamic modes at late times are significantly modified. Rather than continuing with the $1/\tau$ decay as in the gradientless case (shown in the Figure as a dashed line), the hydrodynamic modes obtain a constant real part at late times, as predicted from equations~\eqref{eq:sound-dispersion} and \eqref{eq:shear-dispersion}. The curves associated with the sound modes are shown in red and those associated with the shear mode in blue. The constant eigenvalues at late times for these hydrodynamic modes imply that they will decay exponentially and thus significantly faster than the hydrodynamic modes at zero gradients (which have a power-law decay due to $\epsilon \propto 1/\tau$). This is unsurprising, as hydrodynamic quantities like the momentum $T^{0x}$ are no longer strictly conserved at finite gradients. The exact late-time hydrodynamic eigenvalues depend on the gradient (i.e., the final height of the red and blue curves), as already computed order by order in $k$ in equations~\eqref{eq:sound-dispersion}--\eqref{eq:shear-dispersion}. 

In Figure~\ref{fig:Re-late-vs-k}, we show the numerically computed spectrum at $\tau \rightarrow \infty$ as a function of $k$. As expected, we see the non-hydrodynamic modes in gray at the constant decay rate ($\epsilon_n \sim \tau_R^{-1}$), and the $\propto k^2$ behavior of the sound and shear modes at small gradients. The black dots indicate the poles of the retarded correlators obtained from the exact RTA solution of Ref.~\cite{Romatschke:2015gic}, which perfectly match the sound- and shear-mode dispersion relation of our framework. At larger gradients ($k \tau_R >1$), the expansion of equation~\eqref{eq:sound-dispersion} breaks down and the real parts of the eigenvalues will reach a maximum of $\Re{\epsilon_n} \sim \tau_R^{-1}$. In fact, from the exact solution of the RTA equation, the hydrodynamic pole will no longer exist beyond a critical point $k_c \tau_R\simeq 4.53$~\cite{Romatschke:2015gic}. At this point, there are no longer any hydrodynamic (slow) modes and the adiabatic hydrodynamization framework is no longer useful in analyzing this system. Physically, this represents the gradients driving the system apart on a timescale much shorter than the time the system needs to relax back to thermal equilibrium via collisions,
\begin{align}
\tau_g \equiv \frac{1}{k}\ll\tau_R\,. 
\end{align}

We conclude that transverse spatial gradients qualitatively reorganize the late-time low-lying spectrum. They mix the ground states of the hydrodynamic sectors, producing sound and shear modes that become the slowest excitations of the full system. In the late-time, small-gradient regime, our kinetic-theory description reproduces the hydrodynamic gradient expansion, and we numerically identify the corresponding sound and shear modes. As the wave number increases, the spectral gap between the hydrodynamic mode and the non-hydrodynamic spectrum decreases and eventually closes at $k_c\tau_R \simeq 4.53$. Beyond this point, high-$k$ perturbations are no longer governed by an isolated hydrodynamic mode and do not approach equilibrium along a low-dimensional hydrodynamic attractor. Nevertheless, these modes retain finite damping rates and therefore decay in the absence of continuous external driving. The closing of the gap thus signals the breakdown of a hydrodynamic description of their relaxation, rather than a failure of equilibration itself. Moreover, it does not prevent the bulk of the system as a whole from hydrodynamizing and thermalizing, since the bulk evolution is dominated by low-$k$ modes that remain spectrally separated from non-hydrodynamic excitations. In this sense, high-$k$ modes can equilibrate without hydrodynamizing: their relaxation proceeds through multiple kinetic modes rather than through an isolated hydrodynamic attractor.

\subsection{Sector and moment populations}\label{sec:population-sectors}
\begin{figure}
    \begin{subfigure}{0.49\linewidth}
        \centering
        \includegraphics[width=\linewidth]{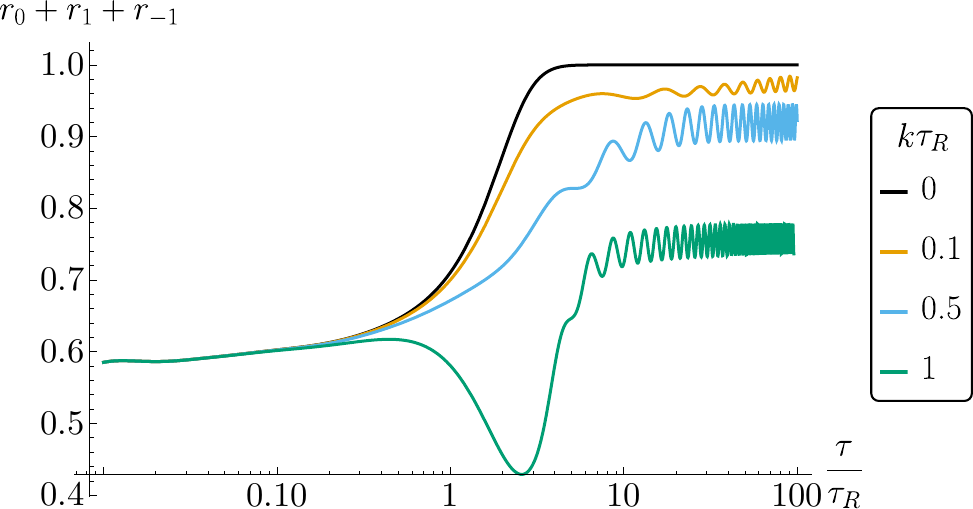}
        \caption{\label{fig:hydro-sector-population-k-finite}}
    \end{subfigure}\hfill
    \begin{subfigure}{0.49\linewidth}
    
        \includegraphics[width=\linewidth]{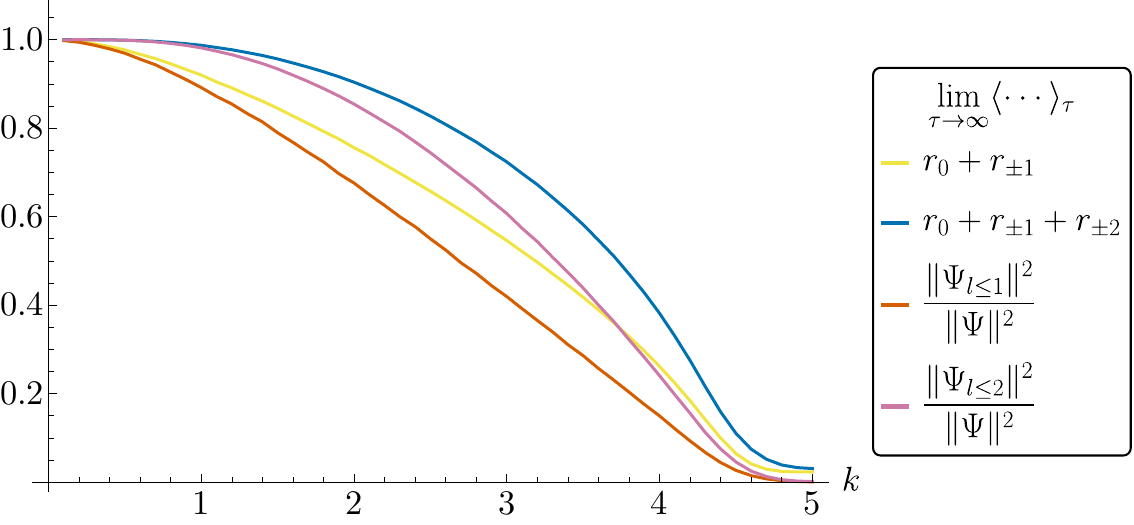}
        \caption{\label{fig:moment-contribution-late-all-data}}
    \end{subfigure}
\caption{\textbf{(a)}: Contribution of the hydrodynamic sectors, i.e., moments with $m = 0, \pm 1$ for a few selected values of $k\tau_R$. While the early-time dynamics is insensitive to the gradients, at late times the contribution of the hydrodynamic sectors can be significantly reduced. Stronger gradients therefore lead to a higher contribution of the non-hydrodynamic sectors. \textbf{(b)}: Time-averaged late-time limit of various measures of hydrodynamic contribution as a function of gradients. We choose the sector contributions $\abs{m} \leq 1, \abs{m}\leq2$ as well as the hydrodynamic moment contributions $l \leq1$ and stress-tensor contributions $l\leq2$ to the distribution as possible measures of the accuracy of hydrodynamics at finite gradients. The dependence can be well fitted to power laws given in the main text.}
\end{figure}

In the previous section we described how the gap between the hydrodynamic and non-hydrodynamic modes closes at large gradients, thereby preventing thermalization via the hydrodynamic attractor path and precluding a reduced hydrodynamic description of the system at large gradients. Here we make this notion more precise by quantifying how well the hydrodynamic description may hold at finite gradients. As a first measure, we may look at the sector population $r_m \equiv \norm{\psi_m}^2/\norm{\Psi}^2$ as defined in Section~\ref{subsec:sector-pop-k0}. The total contribution of all hydrodynamic sectors with $m = 0, \pm1$, for a few gradient values, is shown in Figure~\ref{fig:hydro-sector-population-k-finite}. At zero gradients (black curve), we see that the contribution of the hydrodynamic sectors reaches unity shortly after the relaxation time, as we had already seen in Figure~\ref{fig:sector-population-k0}. That is, at late times the distribution function is perfectly described only by moments with $m=0,\pm 1$. As expected from the qualitative discussion at the beginning of this section, finite gradients do not modify any of the early-time dynamics. However, at late times, the hydrodynamic sector population can be significantly reduced at finite gradients, as shown in the orange, blue, and green curves of Figure~\ref{fig:hydro-sector-population-k-finite}. 

A possibly more intuitive measure is the contribution of the stress-energy tensor components to the full kinetic theory distribution function, as these are the dynamical degrees of freedom described by hydrodynamics. Its components are encoded in the first six moments with $l\leq 2$ of our moment expansion (in the even sector), so that we can similarly measure the contribution of these first moments relative to the full distribution vector $\Psi$, which we write as $\norm{\psi_{l\leq 2}}^2/\norm{\Psi}^2$. We observe qualitatively identical behavior to the sector population from Figure~\ref{fig:hydro-sector-population-k-finite} and therefore conclude that we need moments of both higher $l>2$ and $\abs{m} > 2$ to accurately describe the distribution function. 

In Figure~\ref{fig:moment-contribution-late-all-data} we show the time averaged value of various measures of hydrodynamic contributions in the late-time limit, as a function of the gradient $k$. We choose four possible measures: the contribution of the hydrodynamic sectors $\abs{m} \leq 1$ (yellow), the contribution of all $\abs{m}\leq 2$ moments (blue), the contribution for the hydrodynamic moments with $l\leq1$ corresponding to the conserved energy and momentum densities (red), as well as the contribution of all stress-energy tensor moments with $l\leq 2$ (purple). We clearly see that for all these, the qualitative behavior with respect to the gradient strength is identical: at zero gradients, all curves coincide at unity, reflecting that the distribution function is perfectly described at late times by any partition of moments that includes the three hydrodynamic moments with $l\leq 1$ (which is the case for any of the chosen measures). The contributions then decrease as a power-law in $k$ and reach zero around $k\tau_R \sim 4.5$. This can be fitted with the functional form $y = 1 - a x^b $, with coefficients $(a,b) = \{(0.08,1.59),(0.014,2.70),(0.12,1.43),(0.03,2.38)\}$ for the four curves corresponding to $\{\abs{m} \leq 1, \abs{m} \leq 2, l \leq 1, l\leq 2\}$, respectively. This reinforces and extends the result from the eigenspectrum analysis in Section~\ref{subsec:eigenspectrum-gradients-hydro-limit} and Figure~\ref{fig:Re-late-vs-k} which established that the gap of the system closes around the same value of $k\tau_R\sim 4.5$, removing the distinction between hydrodynamic and non-hydrodynamic moments. In other words, at large gradients all moments of the spherical harmonic expansion are equally important at late times and we no longer see a reduction in the degrees of freedom where the system is well described by just a few moments, i.e., the system no longer thermalizes via a path governed by a single hydrodynamic attractor. 

What this analysis adds to the previous conclusion is the observation that even moderate gradients persistently and significantly increase the late-time contribution of non-hydrodynamic moments with both higher $l$ and higher $m$, so that the system cannot be fully described by hydrodynamics. This result is at least qualitatively consistent to simple Knudsen-number arguments which compare the mean free path to the characteristic length scale of the system. By increasing the gradients, we shorten this characteristic length scale, thereby increasing the Knudsen number and interpolating between hydrodynamic and ballistic regimes.

\begin{figure}
   \begin{subfigure}{0.5\linewidth}
    \centering
    \includegraphics[width=\linewidth]{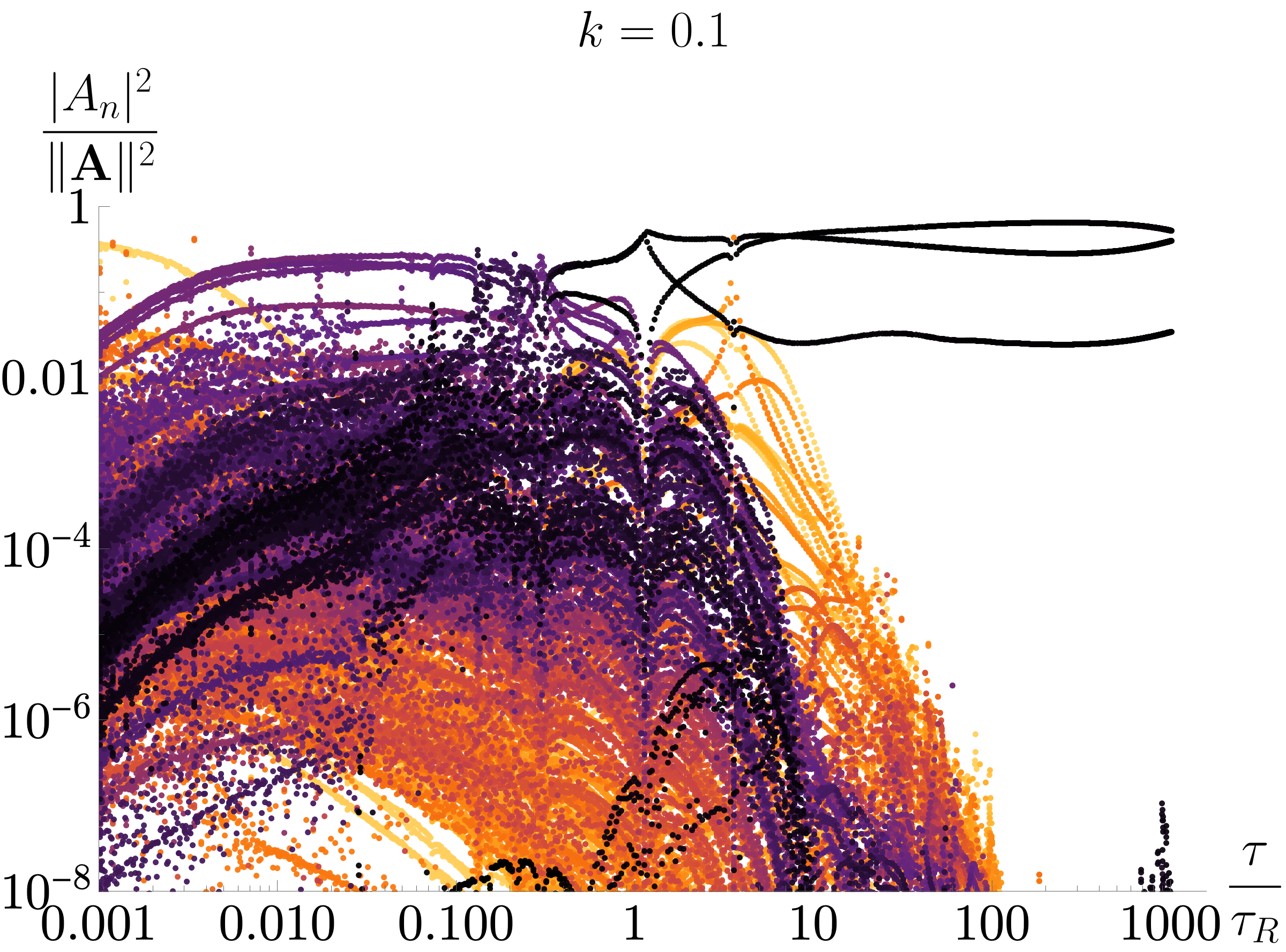}
    \caption{\label{fig:eigendecomp-k0p1-dark}}
   \end{subfigure}
   \begin{subfigure}{0.5\linewidth}
    \centering
    \includegraphics[width=\linewidth]{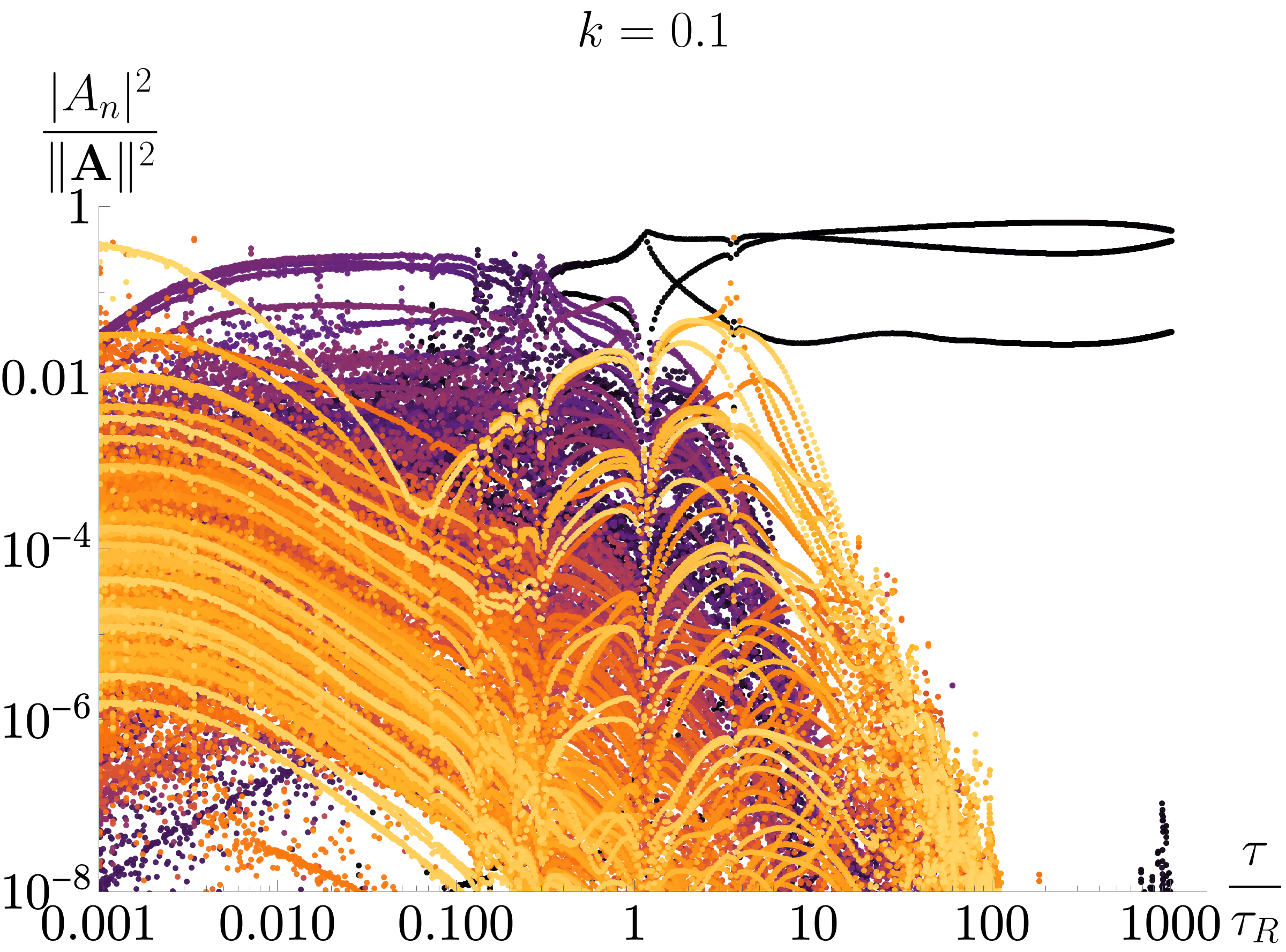}
    \caption{\label{fig:fig:eigendecomp-k0p1-light}}
   \end{subfigure}\\
   
   \begin{subfigure}{0.5\linewidth}
    \centering
    \includegraphics[width=\linewidth]{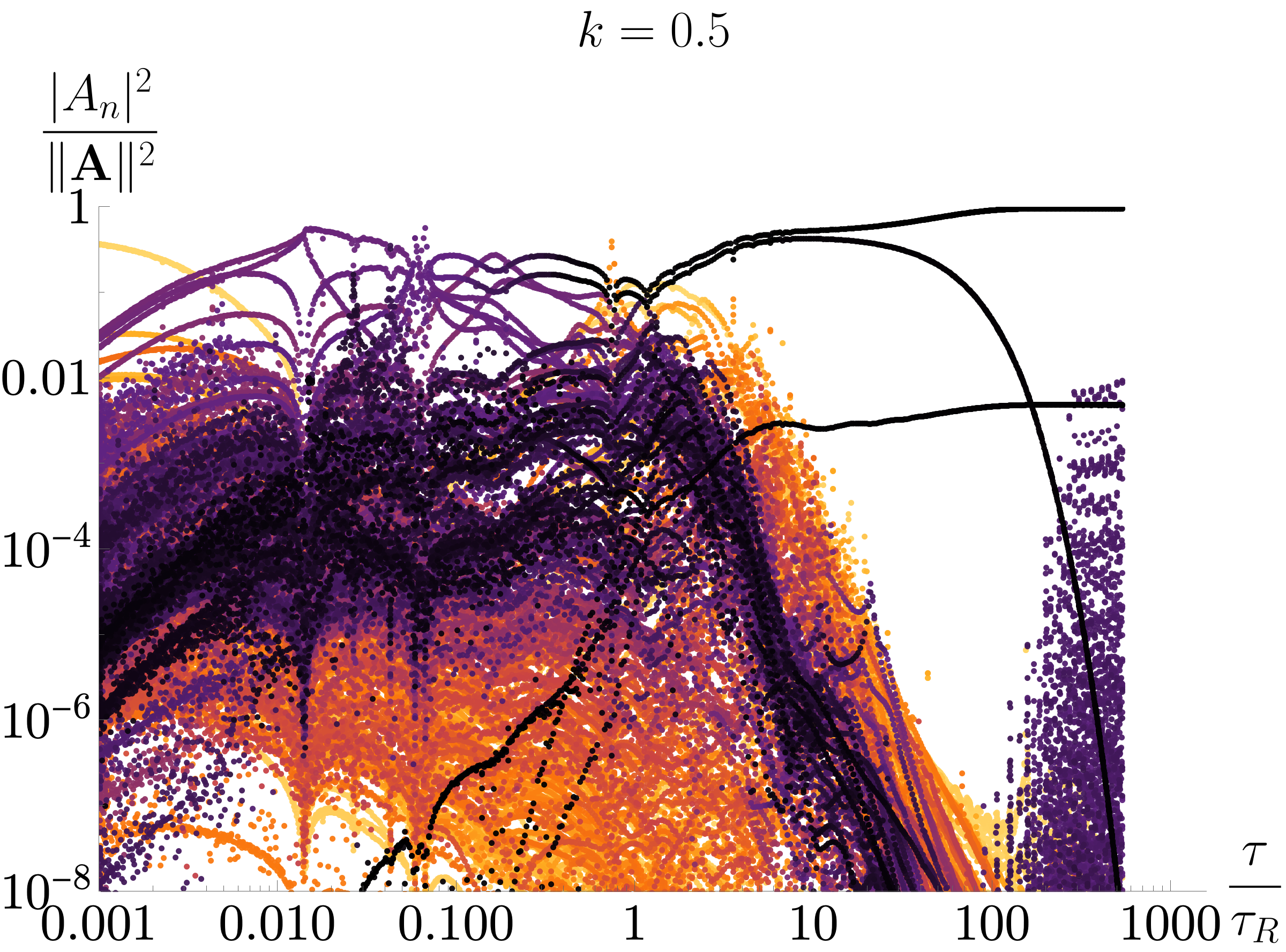}
    \caption{\label{fig:eigendecomp-k0p5-dark}}
   \end{subfigure}
   \begin{subfigure}{0.5\linewidth}
    \centering
    \includegraphics[width=\linewidth]{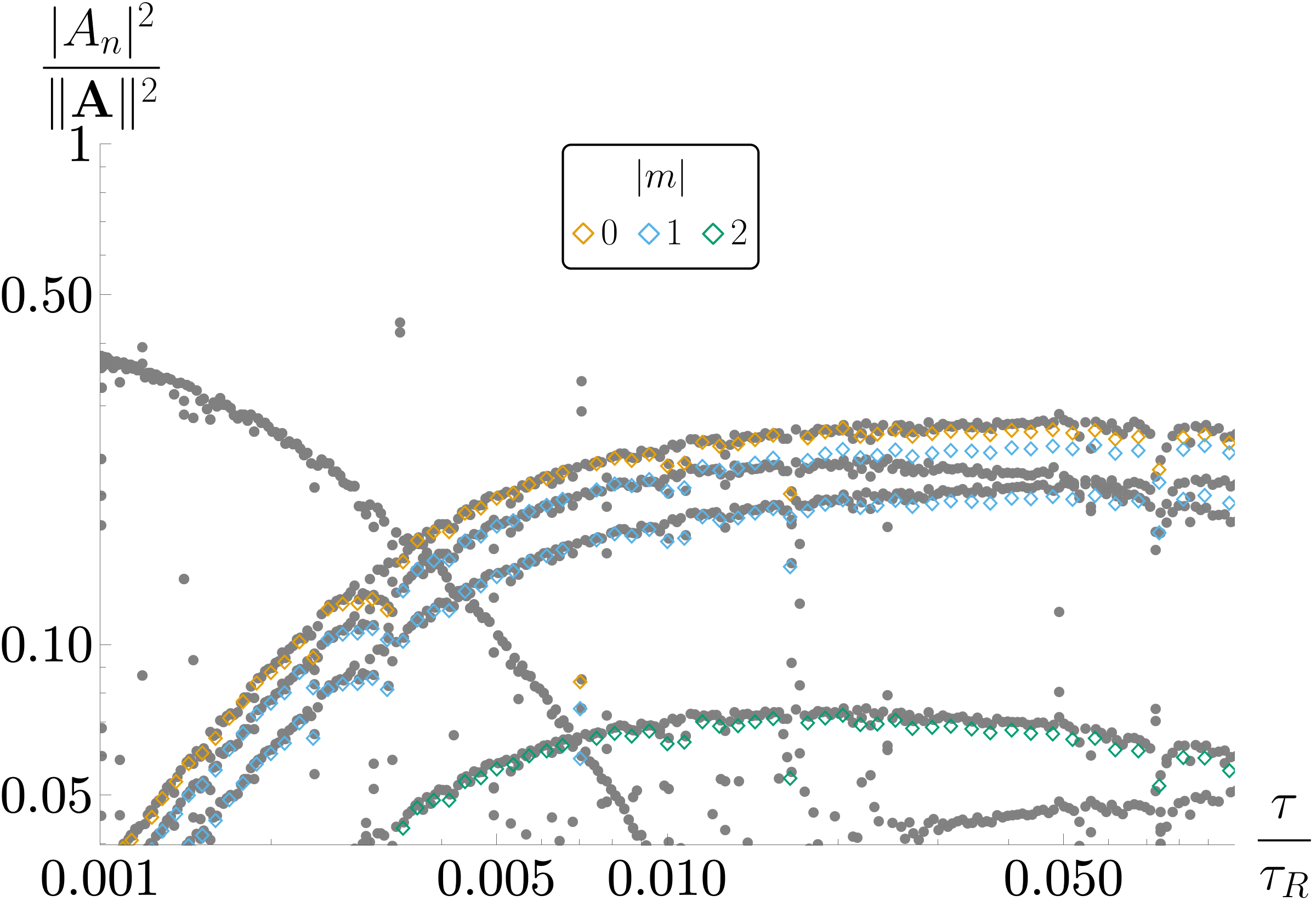}
    \caption{\label{fig:eigendecomp-zoomed}}
   \end{subfigure}
   \caption{\label{fig:full-eigendecomp-k0p1}Coefficients of the full eigenspace decomposition of the even sector moments $\Psi= \sum_n A_n \Phi_n$, normalized by the sum of squared expansion coefficients, at $k \tau_R = 0.1$ (panels \textbf{(a)} and \textbf{(b)}, plotted twice to highlight overlapping regions), and $k \tau_R=0.5$ (panel \textbf{(c)}). We consider the first 231 moments in the spherical harmonic expansion of the even sector, up to $l_\text{max} =  20$. At every timestep the decomposition into the 231 eigenvectors of the coupling matrix is shown, colored consistently according to their decay rates (real part of the eigenvalues), with darker colors indicating smaller eigenvalues. At late times $\tau\gg \tau_R$, the two sound modes and one shear mode found in \eqref{eq:sound-dispersion} and \eqref{eq:shear-dispersion} dominate in the system. At larger gradients (see \textbf{(c)}), the shear mode decays at very late times due to the gap with the sound modes as observed in Figure~\ref{fig:Re-late-vs-k}. At early times, the fast modes (lighter, best seen in panel \textbf{(b)}) are decaying fast, while the slow modes (darker, best seen in panel \textbf{(a)}) gain influence. Panel \textbf{(d)}: zoomed-in part of Figure \textbf{(a)} in gray, with overlaid markers for the decomposition of the full distribution into the decoupled sector ground states of the sectors $\abs{m} = 0, 1, 2$. This explicitly confirms that at early times the system is still decoupled and that each sector is separately dominated by its own ground state, in contrast to the global ground states at late times.}
\end{figure}

\subsection{Transition from sector ground states to global ground states}\label{subsec:sector-gs-to-global-gs}
We now explicitly demonstrate how the system transitions from the decoupled sector ground states at early times to the global ground states, as predicted from our adiabatic hydrodynamization setup. At every timestep, we compute the eigenbasis of the entire system $\{\Phi_n\}$, and compute the coefficients $A_n$ in the decomposition $\Psi = \sum_n A_n \Phi_n$, analogously to Figure~\ref{fig:k=0-eigendecomp}. The results of this computation are shown in Figure~\ref{fig:full-eigendecomp-k0p1}. The coefficients $A_n$ are colored consistently according to the decay rates of their corresponding eigenmodes $\Re{\epsilon_n}$, with darker colors corresponding to smaller eigenvalues (and thus lower decay rates). While the resulting data is somewhat noisy due to the much larger dimensionality of the problem when considering all sectors at once, there are still three interesting observations we can make:
\begin{enumerate}
    \item The fractional occupation of eigenmodes with lower decay rates generally grows (black/purple), whereas the occupation of eigenmodes with higher eigenvalues (yellow) generally decays, highlighting the applicability of the adiabatic hydrodynamization framework at finite gradients.
    \item At early times, the sectors are still approximately decoupled, and so the system is not governed by a single ground state mode, but each sector is well described by its respective instantaneous ground state. We confirm this explicitly in Figure~\ref{fig:eigendecomp-zoomed}, where we computed the overlap of the full distribution with the ground states of the $m = 0, \pm 1, 2$ sectors, and overlaid the results with the full decomposition from Figure~\ref{fig:eigendecomp-k0p1-dark} in gray. The agreement between the gray lines (of the full eigenbasis decomposition) with the colored diamond markers (of the explicit sector ground state decomposition) confirms that even at finite gradients, the same decoupled sector ground states are still reached at early times, and that the same early-time attractors within the sectors remain (see Figures \ref{fig:attractors-k0} and \ref{fig:k=0-eigendecomp}).
    \item At late times, only three ground states remain, which correspond to the mixed-sector sound and shear modes described in Section~\ref{subsec:eigenspectrum-gradients-hydro-limit}. We further observe in Figure~\ref{fig:eigendecomp-k0p5-dark} that at very late times, the shear mode decays with respect to the two sound modes since it has a slightly faster decay rate than the sound modes, as previously shown in Figure~\ref{fig:evals-k-dependence}
\end{enumerate}
We have explicitly shown how the system goes from the early-time, decoupled sector ground state dominated regime, to the late regime containing only three global ground states which correspond to the hydrodynamic modes of the system. The eigenspace decomposition verifies that the adiabatic hydrodynamization framework still holds well in this context with (small) spatial transverse gradients.

\begin{figure}
    \begin{subfigure}{0.5\linewidth}    
        \centering
        \includegraphics[width=\linewidth]{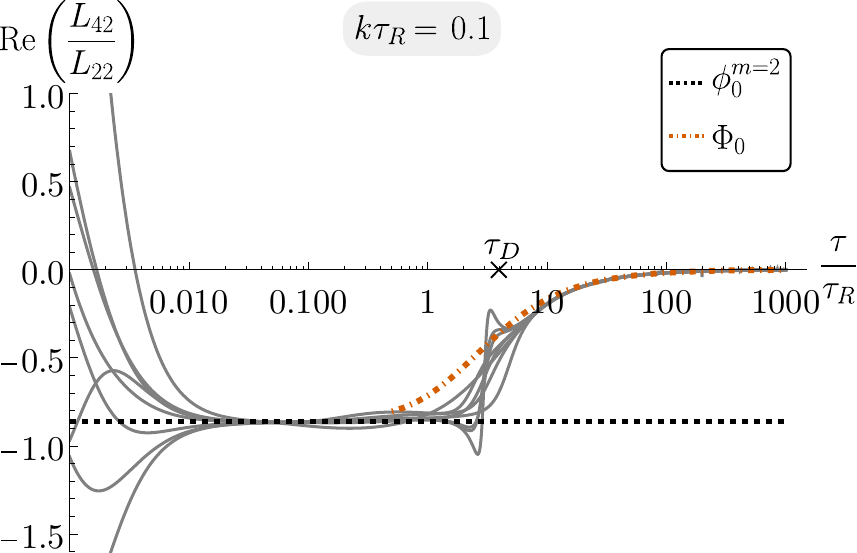}
        \caption{}
        \label{fig:attractor-deviation-vs-time}
    \end{subfigure}
    \begin{subfigure}{0.5\linewidth}    
        \centering
        \includegraphics[width=\linewidth]{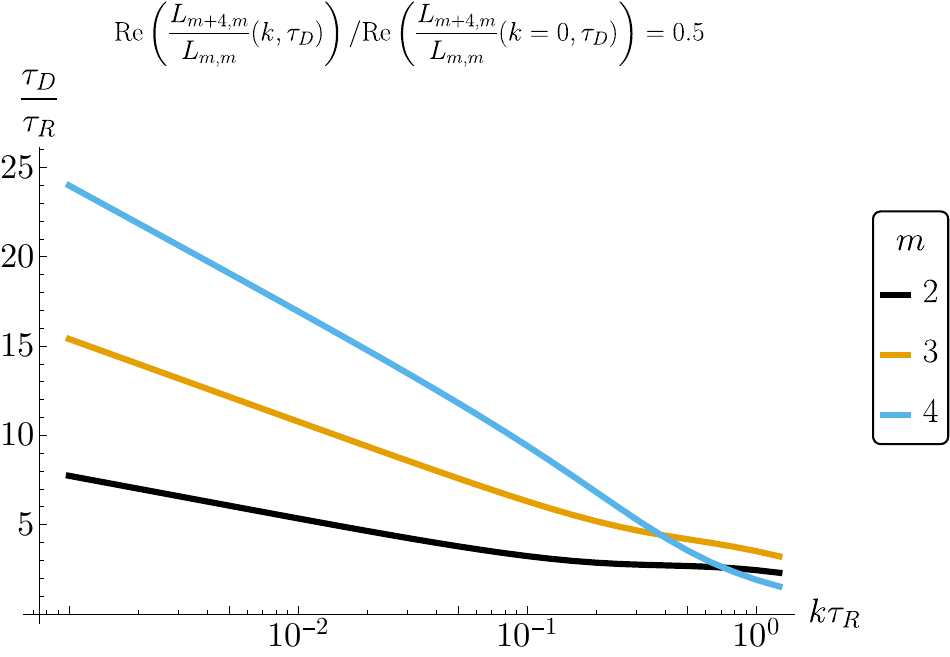}
        \caption{}
        \label{fig:attractor-deviation-vs-k}
    \end{subfigure}
    \caption{\label{fig:attractor-deviation}Modification of the attractors. \textbf{(a)}: Moment ratio $L_{42}/L_{22}$ for different initial conditions at a finite gradient of $k\tau_R=0.1$ (similar to the right plot of Figure~\ref{fig:attractors-k0}). The dashed black curve corresponds to the constant ground state of the $m=2$ sector $\phi_0^{m=2}$, and the orange dash-dotted curve corresponds to the global ground state of the system $\Phi_0$. All the different initial conditions first collapse and decay into the sector ground state at early times, before reaching the global ground state at late times. The deviation time $\tau_D$, defined as the time when the curves have decayed to half the value of the constant ground-state attractor, is further indicated with a cross on the $x$-axis. \textbf{(b)}: Deviation time $\tau_D$ as a function of the gradient, in three different sectors $m=2,3,4$, for the moment ratios $L_{m+4,m}/L_{m,m}$. We observe an approximately logarithmic dependence for small $k$ with different slopes for the sectors, corresponding roughly to a dependence $\exp(\tau_D/\tau_R) \propto (k\tau_R)^{-(m-1)}$. The exact fit values are given in the main text, and additional moment ratios are shown in Appendix~\ref{appendix:deviation-time}.}
\end{figure}

\subsection{Modification of non-hydrodynamic attractors}\label{subsec:modification-attractors}
For the original case at zero transverse gradient, sectors with different $m$ evolve independently and exhibit two qualitatively distinct types of attractor behavior, as discussed in Section~\ref{subsec:no-gradients-recap}. In the non-hydrodynamic sectors, the time-independent eigenbasis causes moment ratios to approach constant, sector-specific attractors. By contrast, moment ratios in the hydrodynamic sectors decay at late times because the dynamics becomes dominated by the conserved hydrodynamic mode. 

Finite transverse gradients couple these otherwise independent sectors and therefore lead to a two-stage evolution. At early times, before the inter-sector coupling becomes important, a moment ratio in a non-hydrodynamic sector first approaches its $k=0$ attractor. At later times, the same ratio departs from this constant value as the global slow mode, which contains components from the hydrodynamic sectors, becomes dominant. The attractor at finite $k$ therefore interpolates between the early-time sector attractor and the late-time global attractor.

This behavior is illustrated in Figure~\ref{fig:attractor-deviation-vs-time} for the ratio $L_{42}/L_{22}$ at $k\tau_R =0.1$, which belongs to the $m=2$ non-hydrodynamic sector. The ratio initially approaches the constant attractor of the decoupled $m=2$ sector, shown by the black dashed line (compare Figure~\ref{fig:attractors-k0}). As the gradient-induced coupling becomes important, it departs from this plateau and approaches the ratio determined by the global ground state, shown by the orange dash-dotted line. To quantify the onset of this crossover, we define the deviation time $\tau_D$ by
\begin{align}
    \frac{L_{42}(k, \tau_D)}{L_{22}(k, \tau_D)}
    =
    \frac{1}{2}
    \left(\frac{L_{42}}{L_{22}}\right)_{k=0,\, \mathrm{attr}}\,.
\end{align}
This point is marked by a cross on the horizontal axis. Increasing $k$ strengthens the inter-sector coupling and therefore decreases $\tau_D$. In contrast, sectors with larger $m$ deviate later because they can couple to the hydrodynamic sectors only through a longer sequence of nearest-neighbor couplings.

We extract $\tau_D$ numerically for the ratios $L_{m+4,m}/L_{m,m}$ and show the results in Figure~\ref{fig:attractor-deviation-vs-k}. Other choices of moment ratios lead to the same qualitative behavior, as demonstrated in Appendix~\ref{appendix:deviation-time}. For $k\tau_R\ll1$, the numerical results are approximately described by
\begin{align}
    \exp\left(\frac{\tau_D}{\tau_R}\right)
    \propto (k\tau_R)^{-a_m},
    \qquad
    a_m=(1.03,\,2.03,\,3.17)
    \quad\text{for}\quad
    m=(2,\,3,\,4),
\end{align}
or, equivalently,
\begin{align}
    \frac{\tau_D}{\tau_R}
    \simeq
    a_m\ln\left(\frac{1}{k\tau_R}\right)+\cdots.
\end{align}
The fitted exponents are close to $a_m=m-1$. This pattern is consistent with the nearest-neighbor structure of the gradient coupling: connecting the hydrodynamic sectors to sector $m$ requires at least $m-1$ successive couplings and hence produces a leading mixing amplitude proportional to $k^{m-1}$. Combining this suppression with the relative growth of the global slow mode naturally gives the observed logarithmic dependence of $\tau_D$ on $k$. Establishing this relation analytically, including its range of validity and subleading corrections, is left for future work.

\section{Conclusions and outlook}
In this work, we have extended the study of far-from-equilibrium attractors to a boost-invariant expanding plasma with transverse spatial gradients. Within the adiabatic hydrodynamization framework, attractor dynamics is governed by the evolution of the slowest instantaneous eigenmodes of the non-Hermitian ``Hamiltonian'' of the kinetic theory. We find that longitudinal expansion and transverse gradients control the evolution on parametrically different timescales. At early times, the rapid longitudinal expansion dominates and suppresses the effects of transverse gradients. The dynamics then approximately decomposes into the independent $m$ sectors of the zero-gradient system, each of which approaches its own sector-specific attractor.

As the system expands, transverse gradients become increasingly important and couple angular moments belonging to different $m$ sectors. The state is therefore no longer confined to the slow mode of an individual sector. Instead, the gradient-induced mixing transfers amplitude between sectors, and the evolution gradually becomes dominated by the globally slowest modes of the coupled system. For sufficiently small $k$, these modes continuously evolve into the hydrodynamic sound and shear modes at late times and span a low-dimensional global attractor manifold. Moment ratios consequently exhibit a two-stage evolution: they first approach their $k=0$ sector attractors and subsequently depart from them toward the global hydrodynamic attractor. The crossover occurs earlier for stronger gradients, whereas sectors with larger $m$ retain their zero-gradient attractor behavior for longer because they couple to the hydrodynamic sectors only through a longer sequence of neighboring-$m$ transitions. At small $k$, the corresponding deviation time depends logarithmically on the gradient scale, with the numerical results consistent with
\begin{align}
    \frac{\tau_D}{\tau_R}
    \sim (m-1)\ln\left(\frac{1}{k\tau_R}\right).
\end{align}

The existence of the late-time hydrodynamic attractor depends crucially on the spectral separation between the hydrodynamic and non-hydrodynamic modes. As $k$ increases, this gap decreases and eventually closes at $k_c\tau_R\simeq4.53$ as is known from Ref.~\cite{Romatschke:2015gic}. Beyond this point $k_c$, the hydrodynamic pole merges with the non-hydrodynamic spectrum, and relaxation is no longer governed by an isolated hydrodynamic mode. High-$k$ perturbations nevertheless retain finite damping rates and decay in the absence of external driving. The gap closure therefore signals relaxation without hydrodynamic-mode dominance, rather than a failure of equilibration. This does not prevent the bulk system from hydrodynamizing, provided that its energy and long-wavelength evolution are dominated by modes with $k<k_c$.

Our results therefore show that transverse gradients do not simply destroy the attractor picture. Instead, they introduce a hierarchy of timescales separating early sector-specific attractors, an intermediate regime of gradient-induced mode mixing, and, at sufficiently small $k$, a late-time hydrodynamic attractor manifold. This establishes more precisely the range of spatial scales over which attractor dynamics can provide an effective description of the far-from-equilibrium plasma and constitutes a necessary step toward assessing its phenomenological relevance in heavy-ion collisions.

Several important extensions remain. First, although individual transverse Fourier modes provide a clean setting in which to isolate the effects of gradients, realistic heavy-ion collisions contain a broad spectrum of spatial fluctuations whose evolution can become coupled beyond linear response. Extending the present analysis to real-space dynamics, including nonlinear coupling between different wave numbers, would provide a more direct connection to phenomenology. Second, relaxing boost invariance would introduce rapidity-dependent fluctuations and generally mix sectors distinguished here by their parity under $p_z\leftrightarrow-p_z$, adding another level of mode coupling to the attractor dynamics. Finally, recent studies have incorporated realistic elastic and inelastic QCD collision kernels into the adiabatic hydrodynamization framework~\cite{Rajagopal:2025nca}. Including transverse gradients within such microscopic kinetic descriptions is an important direction for future work.

\section*{Acknowledgments}
We would like to thank Gabriel Denicol, Yonadav Barry Ginat, Krishna Rajagopal, Bruno Scheihing-Hitschfeld, Rachel Steinhorst, Giorgio Torrieri, Li Yan, Yi Yin, and Fabian Zhou for helpful discussions. This research was partially supported by a University Research Fellowship from the Royal Society under grant URF\textbackslash R1\textbackslash241231, by the Leverhulme Trust under grant LIP2020-014, and by the DFG (German Research Foundation) -- Project-ID 273811115 -- SFB 1225 ISOQUANT. The work of WK is supported by the National Natural Science Foundation of China under Grant No.\ 12575140.

\section*{Use of generative AI and AI-assisted technologies}
The authors used ChatGPT 5.6 Sol to generate initial draft versions of the analytic arguments presented in Appendices~\ref{appendix:real-part-eval-bounds} and \ref{appendix:trunc-dependence-gap}. The authors reviewed, checked, and edited the content as needed and take full responsibility for the content of the publication.

\bibliographystyle{JHEP}
\bibliography{references-zotero,references}

\appendix
\section{Appendix}
\subsection{Boltzmann equation for boost-invariant plasmas}\label{appendix:derivation-boltzmann}
The Boltzmann equation for a relativistic single-particle distribution on the mass shell $f = f(x^\mu, p^i)$, in the absence of external forces, is given by \begin{align}
p^\mu \partial_\mu f + G^i \partial_{p^i}f= -C[f],
\end{align}
where $G^\mu = - \Gamma^\mu_{\lambda\nu} p^\lambda p^\nu $, $\Gamma$ are the standard Christoffel symbols, and $C[f]$ is the collision integral. We define Milne coordinates $x^\mu = (\tau,x,y,\eta)$, with the metric $g_{\mu\nu} = \diag(-1,1,1,\tau^2)$, which are related to the standard Minkowski coordinates $(t,x,y,z)$ by $t = \tau \cosh\eta, z = \tau \sinh\eta$. The Milne metric is Ricci-flat and describes a rapidly expanding spacetime in one direction ($v_z = z/t$ in natural units with $c=1$). The only nonzero Christoffel symbols are $\Gamma^\tau_{\eta \eta} = \tau, \Gamma^\eta_{\eta \tau} = 1/\tau$, such that
\begin{align}
\frac{1}{p^\tau} \left(p^\mu\pdv{f}{x^\mu} + G^i \pdv{f}{p^i} \right)=  \pdv{f}{\tau} + \frac{p^x}{p}\pdv{f}{x} + \frac{p^y}{p} \pdv{f}{y} + \frac{p^\eta}{p} \pdv{f}{\eta} - \frac{2}{\tau} p^\eta \pdv{f}{p^\eta}.
\end{align}
We now assume boost invariance, $\partial_\eta f = 0$, such that we can freely choose $\eta =z = 0$, and then change back to Minkowski coordinates but with proper time $(\tau,x,y,z)$
\begin{align}
\left(\pdv{f}{\tau}\right)_{\eta,p^\eta} = \left(\pdv{f}{\tau}\right)_{z,p^z} + \frac{p^z}{\tau} \left(\pdv{f}{p^z}\right)_{\tau,z},
\end{align}
where we have also used that $p^\tau = p^t, p^z = \tau p^\eta, t = \tau, (\partial z/\partial\tau)_{\eta,p^\eta} =0$ at $z=0$. Substituting this relationship and Fourier transforming in $(x,y)$, we arrive at the kinetic equation for boost-invariant plasmas
\begin{align}
   \pdv{f}{\tau} + i \frac{\vec{k}_\perp \cdot \vec{p}_\perp}{p}f - \frac{p^z}{\tau}  \pdv{f}{p^z} =-C[f].
\end{align}

\subsection{Derivation of the moment equations}\label{appendix:derivation-moment-eqns}
We consider the energy-density weighted distribution $F(\theta,\phi) \equiv \frac{1}{2\pi^2} \int \dd{p} 
p^3 f$. Substituting into the kinetic equation in spherical momentum space coordinates $(p,\theta,\phi)$, with $\cos\theta= p_z/p, \tan\phi= p_y/p_x$, and setting $\tau_R = 1$, we have
\begin{align}
    \partial_\tau F &= \frac{1}{\tau} \left[-4 \cos^2 \theta + \sin^2 \theta \cos\theta \pdv{}{\cos{\theta}}\right]F - \left(F-F_\text{eq}\right) \\
    &- i \left[\sin\theta \left(k_x \cos\phi + k_y \sin\phi\right) \right] F\\
    \partial_\tau F &= -\frac{1}{\tau} \hat{H}_{\text{free}} F - \hat{C}[F] - i \hat{H}_{g} F,
\end{align}
where in the final line we have written the equation in terms of a free-streaming part ($\hat{H}_\text{free}$), a gradient part ($\hat{H}_{g}(\vec{k})$), and a collision integral ($\hat{C}[F]$).
Consider the spherical harmonic functions defined by\footnote{Note that these differ by a factor of $\sqrt{4\pi}$ from the usual textbook definition.}
\begin{align}\label{eq:Ylm-definition}
    Y_l^m(\theta,\phi) \equiv (n_l^m)^{-1} P_l^m(\cos\theta) e^{i m \phi}, \quad   n_{l}^{m}  = \sqrt{\frac{1}{2l+1}\frac{(l+m)!}{(l-m)!}},
\end{align}
where $P_l^m$ are the standard associated Legendre polynomials. The spherical harmonics obey an orthonormality relation with respect to the average over the solid angle \begin{align}
    \langle Y_l^m,Y_{l'}^{m'} \rangle \equiv \int (Y_{l}^{m})^* Y_{l'}^{m'} \frac{\dd{\Omega}}{4\pi} = \frac{(-1)^m}{4\pi} \int_0^\pi \int_{0}^{2\pi} Y_{l}^{-m} Y_{l'}^{m'} \sin\theta \dd{\theta}\dd{\phi} = \delta_{l,l'}\delta_{m,m'}
    \label{eq:inner-product}
\end{align}
The spherical harmonics further obey the following recurrence relations:
\begin{align}
    \cos\theta Y_l^m &= a_l^m Y_{l-1}^{m} + a_{l+1}^{m} Y_{l+1}^m, \quad
    \cos^2\theta Y_l^m = b_{l}^{m} Y_{l-2}^{m} + c_{l}^{m} Y_{l}^{m} + b_{l+2}^{m} Y_{l+2}^{m}\\
    \sin\theta Y_l^m &= \frac{e^{-i \phi}}{(2l+1)} \left(\frac{n_{l-1}^{m+1}}{n_{l}^{m}} Y_{l-1}^{m+1} - \frac{n_{l+1}^{m+1}}{n_{l}^{m}} Y_{l+1}^{m+1}  \right)\\
    \pdv{Y_l^m}{\cos\theta} &= -\frac{1}{\sin\theta} \left( e^{-i \phi} \frac{n_{l}^{m+1}}{n_{l}^{m}} Y_l^{m+1} + m \cot\theta Y_{l}^{m} \right)
\end{align}
with the numerical coefficients
\begin{align}
    (a_l^m)^2 = \frac{(l-m)(l+m)}{(2l+1)(2l-1)}, \quad b_{l}^{m} = a_l^m a_{l-1}^{m}, \quad c_{l}^{m} = (a_l^m)^2 + (a_{l+1}^{m})^2
 \end{align}
We can then compute the coupling in this basis:
\begin{align}
    \langle Y_l^m, \hat{H}_\text{free} Y_{l'}^{m'} \rangle &= \langle Y_l^m, \left[4 \cos^2 \theta - \sin^2 \theta \cos\theta \pdv{}{\cos{\theta}} \right] Y_{l'}^{m'} \rangle \\
   &=  (4+m)\langle Y_l^m, \cos^2 \theta Y_{l'}^{m'} \rangle + \frac{n_{l'}^{m'+1}}{n_{l'}^{m'}}\langle Y_l^m, \sin \theta \cos\theta e^{-i \phi} Y_{l'}^{m'+1}  \rangle \\
   &=  \delta_{m,m'} \left( A_{l}^{m} \delta_{l',l} + B_{l}^{m} \delta_{l',l-2} + C_{l}^{m} \delta_{l',l+2} \right)
   \end{align}
with the coefficients given by
\begin{align}
   A_{l}^{m} &= (4+m) c_{l}^{m} + \frac{1}{2l+1} \frac{n_{l}^{m+1}}{(n_{l}^{m})^2} \left(n_{l-1}^{m+1} a_{l}^{m+1} - n_{l+1}^{m+1} a_{l+1}^{m+1} \right),\\
   B_{l}^{m} &= (4+m) b_{l}^{m} + \frac{a_{l-1}^{m+1}}{2l+1} \frac{n_{l-2}^{m+1} n_{l-1}^{m+1} }{n_{l-2}^{m} n_{l}^{m}},\\
   C_{l}^{m} &= (4+m) b_{l+2}^{m} - \frac{a_{l+2}^{m+1}}{2l+1} \frac{n_{l+2}^{m+1} n_{l+1}^{m+1} }{n_{l+2}^{m} n_{l}^{m}} = B_{l+2}^m \left(\frac{5}{4+l}-1\right).
\end{align}
These coefficients are the same as already calculated in Ref.~\cite{brewerFarfromequilibriumSlowModes2022}, and for the case $m=0$ in Ref.~\cite{blaizot_fluid_2018}, but rescaled with a factor of $n_{l'}^{m'}/n_{l}^{m}$ due to a different normalization of the spherical harmonics. For the transverse parts, consider
\begin{align}
\frac{\vec{k}_\perp\cdot \vec{p}_\perp}{p} = k_x \sin\theta \cos\phi + k_y \sin\theta \sin\phi  =  \frac{1}{\sqrt{6}} \left( (k_x + i k_y) Y_1^{-1} - (k_x-i k_y) Y_1^1 \right)
\end{align}
The relevant overlap integral to consider is therefore
\begin{align}
    \langle Y_l^m, Y_1^{1} Y_{l'}^{m'} \rangle &= \int (Y_l^m)^* Y_1^{1} Y_{l'}^{m'} \frac{\dd{\Omega}}{4 \pi} \\
    &= (-1)^m \sqrt{3(2l+1)(2l'+1)} \begin{pmatrix}
        l & 1 & l' \\
        0 & 0 & 0
    \end{pmatrix}\begin{pmatrix}
        l & 1 & l' \\
        -m & 1 & m'
    \end{pmatrix}
\end{align}
where we have used the Wigner 3j symbols and the fact that $(Y_l^m)^* = (-1)^m Y_l^{-m}$. The selection rules for the Wigner symbols imply that this integral is zero unless $m' = m-1$ and $l' = l \pm 1$.
Repeating analogous steps for $Y_1^{-1}$ and extracting the coefficients, we obtain
\begin{align}
    \begin{split}
    \langle Y_l^m, (k_x \sin\theta \cos\phi + k_y \sin\theta\sin\phi) Y_{l'}^{m'} \rangle = \delta_{m',m-1} (k_x - i k_y) \left(D_{l'}^{-m'} \delta_{l',l+1} - D_l^m \delta_{l',l-1} \right) \\ 
    +\delta_{m',m+1} (k_x + i k_y) \left(  D_l^{-m} \delta_{l',l-1}-D_{l'}^{m'}\delta_{l',l+1}\right).
    \end{split}
\end{align}
Since we choose $C[F]= \tau_R^{-1}(F- \sum_{(lm)_H} L_{lm} Y_l^m)$, it follows immediately that 
\begin{align}
    C= \delta_{m,m'}\delta_{l,l'} (1- \delta_{l,0})(1- \delta_{l,1})/\tau_R
\end{align}
We summarize and give explicit expressions for the coefficients, writing $O \equiv \langle Y_l^m, \hat{O} Y_{l'}^{m'} \rangle$ as a shorthand for the relevant operators.
\begin{align}
    H_\text{free} &=  \delta_{m,m'} \left( A_{l}^{m} \delta_{l',l} + B_{l}^{m} \delta_{l',l-2} + C_{l}^{m} \delta_{l',l+2} \right) \label{eq:H-free-app} \\[10pt]
    \begin{split}
    H_g &= \delta_{m',m-1} (k_x - i k_y) \left( D_{l'}^{-m'} \delta_{l',l+1} - D_l^m \delta_{l',l-1}\right)\\
    & + \delta_{m',m+1}(k_x + i k_y) \left(D_l^{-m} \delta_{l',l-1}-D_{l'}^{m'}\delta_{l',l+1} \right)
    \end{split}\\[10pt]
  C &= \delta_{m,m'}\delta_{l,l'} (1- \delta_{l,0})(1- \delta_{l,1})/\tau_R
\end{align}
with explicit expressions for the coefficients given by

\begin{align}
A_l^m &=    \frac{7 l^2+7 l-5 m^2-4}{(2 l-1) (2 l+3)},&\quad B_l^m &= \frac{l+2}{2 l-1} \sqrt{\frac{((l-1)^2-m^2) (l^2-m^2)}{(2 l-3) (2 l+1)}}, \nonumber\\
C_l^m &= B_{l+2}^m \left(\frac{5}{4+l}-1\right),  &\quad D_l^m &= \frac{1}{2} \sqrt{\frac{(l+m-1) (l+m)}{(2l+1)(2l-1)}}. \label{eq:ABCD}
\end{align}

\subsection{Relation to the stress-energy tensor}\label{appendix:stress-tensor-relations}
The stress-energy tensor in a conformal kinetic theory is given by 
\begin{align}
    T^{\mu\nu} = \int \frac{\dd^3{\vectorbold{p}}}{(2\pi)^3 p} p^\mu p^\nu f = \frac{1}{2} \int_{-1}^{1} \dd{\cos\theta} \int_{0}^{2\pi} \frac{\dd{\phi}}{2\pi}  \frac{1}{2\pi^2} \int_{0}^{\infty}\dd{p} p \cdot p^\mu p^\nu f =\left\langle \frac{p^\mu p^\nu}{p^2}, F \right\rangle.
\end{align}
Using the definition $L_{lm} = \langle Y_l^m, F\rangle$ we immediately get the relations to the stress-energy tensor components:
\begin{align}
\begin{split}
    &L_{00} =T^{00}, \quad L_{10}=\sqrt{3} T^{0z} ,\quad  L_{1,\pm1}=\frac{\sqrt{6}}{2} \left(\mp T^{0x}+i T^{0y}\right)\\
    &L_{2,\pm1}=\frac{\sqrt{30}}{2} \left(\mp T^{xz}+i T^{yz}\right), \quad L_{2,\pm 2}=\frac{\sqrt{30}}{4} \left(\mp 2 i  T^{xy}+ T^{xx}- T^{yy}\right),\\
    &L_{20}=\frac{\sqrt{5}}{2} \left(3T^{zz}-  T^{00}\right)
\end{split}
\end{align}

\subsection{Methods}\label{appendix:methods}
Random initial conditions were chosen in solving equation \eqref{eq:coupling-matrix}, although we note that none of the results were sensitive to the initial conditions except where explicitly stated. To minimize edge effects, we scaled the initial conditions such that the absolute value of the moments within the vector $\Psi = \{L_{lm}\}$ decreased exponentially with increasing $l$. For the computation with gradients, a truncation of $l_\text{max} = 50$ was chosen, corresponding to the first 1326 moments in $\Psi$. For the zero-gradient computation, a truncation of $l_\text{max} = 500$ was used, corresponding to the first 251 moments of each sector vector.

\subsection{Attractor deviation time}\label{appendix:deviation-time}
\begin{figure}
\begin{subfigure}{0.5\textwidth}
    \centering
    \includegraphics[width=\linewidth]{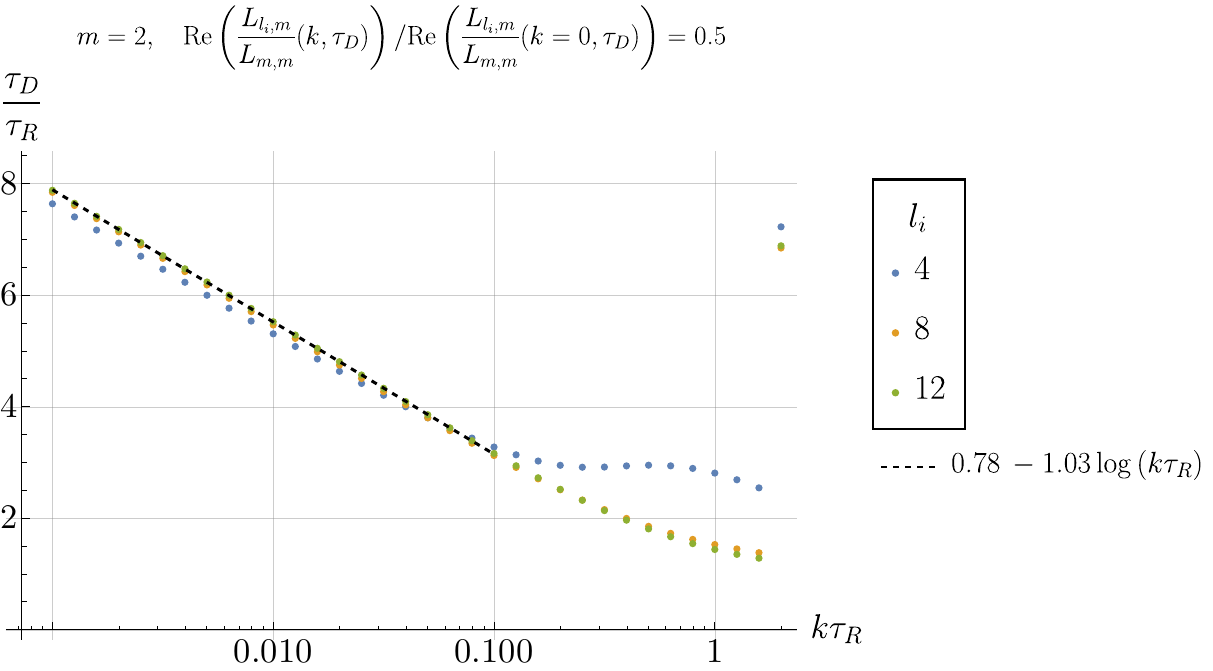}
    \caption{}
\end{subfigure}
\begin{subfigure}{0.5\textwidth}
    \centering
    \includegraphics[width=\linewidth]{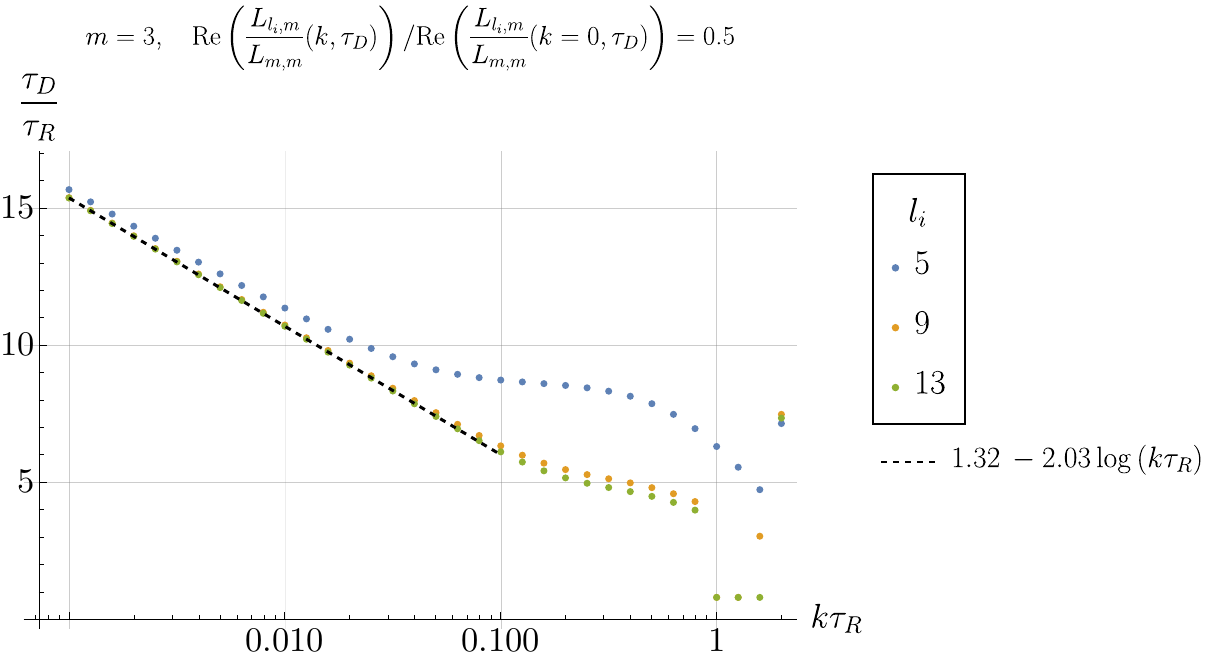}
    \caption{}
\end{subfigure}\\
\begin{subfigure}{0.5\textwidth}
    \centering
    \includegraphics[width=\linewidth]{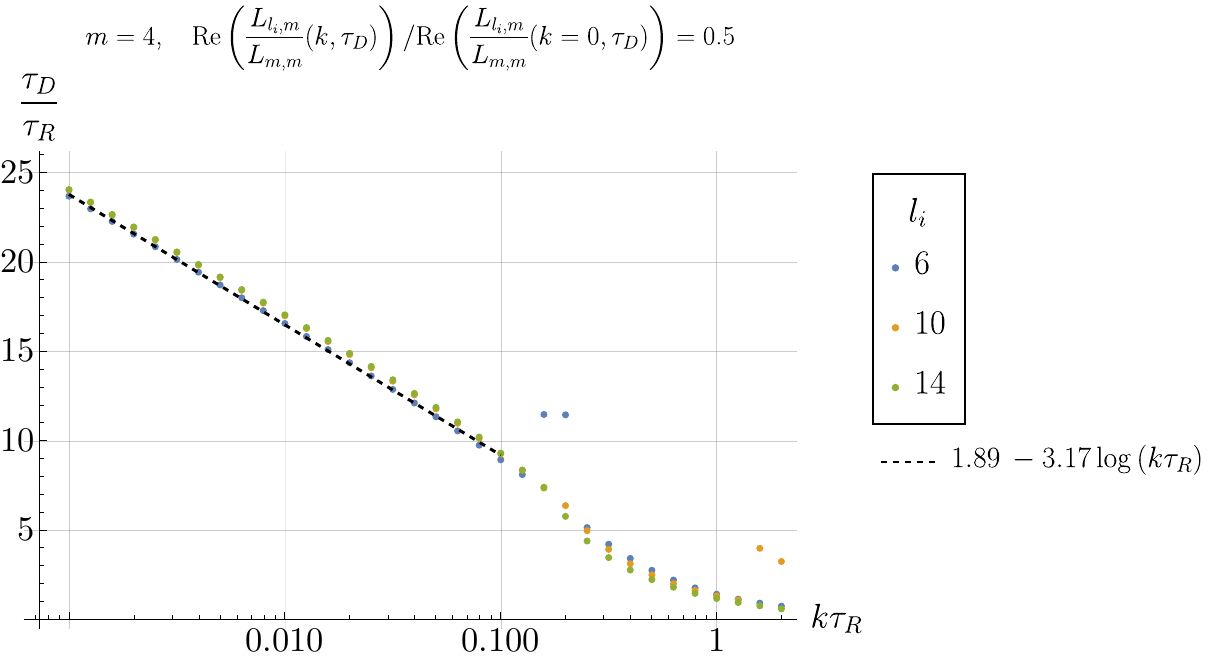}
    \caption{}
\end{subfigure}

 \caption{\label{fig:deviation-attractor}Time at which the finite gradient solution deviates from the non-hydrodynamic attractor for various sectors $(m,+)$. The attractor deviates at a time $\exp(\tau_D/\tau_R) \propto (k\tau_R)^{-a}$ for small $k \tau_R\lesssim 0.1$, and with $a \approx 1,2,3$ for $m=2,3,4$, respectively.}
\end{figure}

In Figure \ref{fig:deviation-attractor} we display additional data to complement Figure \ref{fig:attractor-deviation}. We specifically note that the dependence on gradients is insensitive to the exact moment ratio chosen. The exact fits are given in the figure legends and seem to roughly obey $\exp(\tau_D/\tau_R) \propto (k\tau_R)^{1-m}$.

\subsection{Bounds on the real parts of the eigenvalues of $H_\text{free}$ for $m\geq2$}\label{appendix:real-part-eval-bounds}
The basic idea of the following argument is to use a diagonal similarity transform to bring the coupling matrix into a form whose off-diagonal part is antisymmetric, so that the real parts of its eigenvalues are bounded by the diagonal entries alone.

Consider the free, tridiagonal coupling matrix \eqref{eq:H-free-app} at a fixed value of $m$, which we call $H^{(m)}$, truncated at some finite $l\leq L$; all steps below hold for every $L$ and the resulting bounds are uniform in $L$. It decomposes into two independent parity sectors ($s = \pm 1= (-1)^{l+m}$), with index set $\mathcal{I}=\{l_0,l_0+2,\dots,L\}$ and $l_0 = m + \delta_{s,-1}$ the lowest $l$ value within the sector. Neighboring elements of $H^{(m)}$, $l$ and $l+2$, are connected by
\begin{align}
H_{l,l+2}^{(m)}
=
C_l^m
=
-r_lB_{l+2}^m = -r_l H_{l+2,l}^{(m)}
\end{align}
with $r_l
\equiv
\frac{l-1}{l+4} \in (0,1)$ for $l\geq 2$. 
We therefore construct $\widetilde H^{(m)}=P^{1/2}H^{(m)}P^{-1/2}$, where the diagonal matrix $P$ follows recursively from $p_{l+2}/p_l=r_l$,
\begin{align}
P_{ll'}&=p_l\delta_{ll'},  \quad
p_{l_0+2n}=
p_{l_0}
\prod_{j=0}^{n-1}
\frac{l_0+2j-1}{l_0+2j+4}
\end{align}
for $l=l_0+2n$. Since $m\geq2$ we have $l_0\geq2$ and hence $r_l>0$ throughout the sector, so all $p_l>0$ and $P^{\pm1/2}$ are well defined. The transformed upper/lower off-diagonal element is then
\begin{align}
\begin{split}\label{eq:Htilde}
\widetilde H_{l,l+2}^{(m)}&=
\sqrt{\frac{p_l}{p_{l+2}}}\,C_l^m
=
\frac{1}{\sqrt{r_l}}
\left(-r_lB_{l+2}^m\right)
=
-\sqrt{r_l}\,B_{l+2}^m,\,\\
\widetilde H_{l+2,l}^{(m)}&=
\sqrt{\frac{p_{l+2}}{p_l}}\,B_{l+2}^m=\sqrt{r_l}\,B_{l+2}^m.
\end{split}
\end{align}
Thus, $\widetilde H_{l,l+2}^{(m)}=-\widetilde H_{l+2,l}^{(m)}$, while the diagonal is unchanged by the transform, $\widetilde H_{ll}^{(m)}=A_l^m$, and the eigenvalues are those of $H^{(m)}$.

Now, let $q_n$ be a right eigenvector $\widetilde H^{(m)}q_n =
\lambda_n^{(m)}q_n$.
Multiplying by $q_n^\dagger$ from the left and taking the real part gives
\begin{align}
\operatorname{Re}\lambda_n^{(m)}
&=
\frac{1}{2}\frac{q_n^\dagger\left[\widetilde H^{(m)}+\widetilde H^{(m)\dagger}\right]q_n}{q_n^\dagger q_n} =
\frac{
\displaystyle\sum_{l\in \mathcal{I}}
A_l^m|q_{n,l}|^2
}{
\displaystyle\sum_{l\in \mathcal{I}}
|q_{n,l}|^2
}.
\label{eq:real-eigenvalue-average}
\end{align}
Hence $\operatorname{Re}\{\lambda_n^{(m)}\}$ is a convex combination of the diagonal coefficients $A_l^m$. In particular,
\begin{align}
\min_{l\in\mathcal I}A_l^m
\leq
\operatorname{Re}\lambda_n^{(m)}
\leq
\max_{l\in\mathcal I}A_l^m.
\label{eq:A-eigenvalue-bound}
\end{align}
Equality in \eqref{eq:A-eigenvalue-bound} would require $q_n$ to be supported on a single index $l$, which is impossible for a tridiagonal matrix with $B_{l+2}^m\neq0$; both inequalities are therefore strict.

It remains to bound the $A_l^m$ themselves. Using $(2l-1)(2l+3)=4l^2+4l-3$, the diagonal coefficients can be written as
\begin{align}
A_l^m = \frac{7l^2+7l-5m^2-4}
{(2l-1)(2l+3)} = \frac{7}{4}-\frac{5\left(4m^2-1\right)}{4(2l-1)(2l+3)}.
\end{align}
For $m\geq 2$ the numerator $5(4m^2-1)$ is positive, and $(2l-1)(2l+3)$ is positive and increasing in $l$ for $l\geq l_0\geq 2$. Hence $A_l^m<7/4$ for any $l$, $A_l^m$ is monotonically increasing in $l$, and $\lim_{l\to\infty}A_l^m = 7/4$. This already demonstrates the upper bound on the real parts of the eigenvalues. The minimum is attained at $l=l_0$ and distinguishes the two sectors: for the even $s=+$ sector, $l_0=m$ and
\begin{align}
\min_{l\in\mathcal I}A_l^m = A_{m}^m = \frac{7}{4}-\frac{5(2m+1)}{4(2m+3)}=\frac{m+4}{2m+3}>\frac{1}{2},
\end{align}
while for the odd $s = -$ sector, $l_0=m+1$ and
\begin{align}
\min_{l\in\mathcal I}A_l^m = A_{m+1}^m = \frac{7}{4}-\frac{5(2m-1)}{4(2m+5)}=\frac{m+10}{2m+5}>\frac{1}{2}.
\end{align}
Note that the restriction to $m\geq2$ enters through $l_0\geq2$: for $\abs{m}=1$ in the $s=+$ sector one has $l_0=1$ and $r_1=0$, so that $P$ becomes singular and the construction degenerates, while for $m=0$ one finds $A_l^0=\frac{7}{4}+\frac{5}{4(2l-1)(2l+3)}>\frac{7}{4}$ and the upper bound fails.

In summary, for $m\geq2$,
\begin{align}
\begin{cases}
\frac{1}{2}<\frac{m+4}{2m+3}<\operatorname{Re}\lambda_n^{(m)} < \frac{7}{4} & \textrm{for even sectors},\\
\frac{1}{2}<\frac{m+10}{2m+5}<\operatorname{Re}\lambda_n^{(m)} < \frac{7}{4} & \textrm{for odd sectors.}
\end{cases}
\end{align}

\subsection{Truncation dependence of the gap}\label{appendix:trunc-dependence-gap}
\begin{figure}
    \centering\includegraphics[width=.9\linewidth]{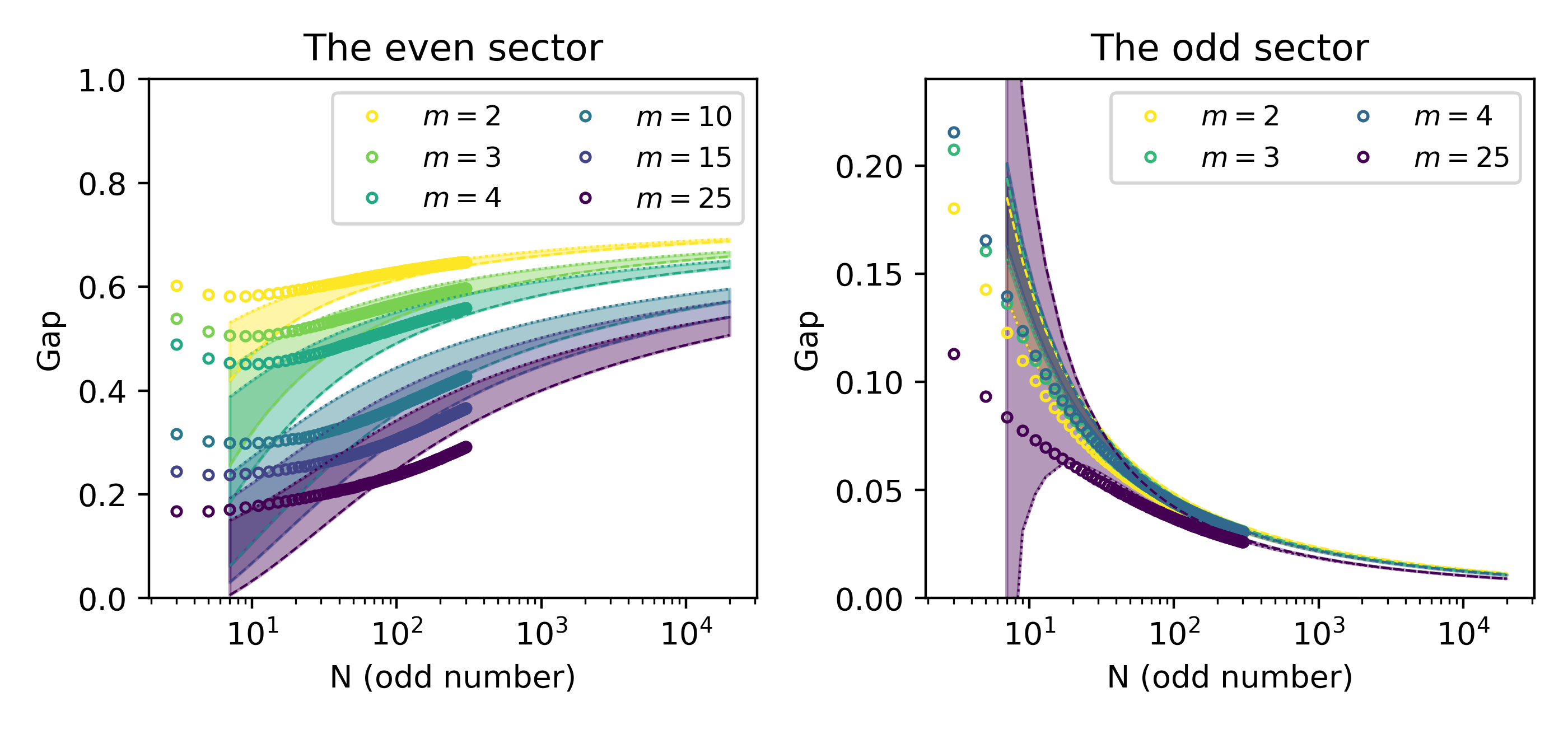}
\caption{Comparison of the finite-$N$ spectral gaps with their asymptotic behavior. The left panel shows the even-sector gap for different values of $m$. The circles are obtained by numerical diagonalization of the truncated Hamiltonian, the dotted curves show Eq.~\eqref{eq:even-gap}, and the dashed curves include one Newton iteration using Eq.~\eqref{eq:even-gap} as the initial guess. The right panel shows the corresponding comparison for the odd sector with odd $N$. For even $N$, the two lowest odd-sector eigenvalues have identical real parts. In both sectors, the gaps approach their continuum limits, $\Delta_\infty^{(+)}=3/4$ and $\Delta_\infty^{(-)}=0$, only logarithmically with $N$.}
    \label{fig:Gap-vs-N-freestream}
\end{figure}

We study the spectral gap of the free-streaming Hamiltonian and its
approach to the continuum limit as the number of retained moments
$N$ increases. The basic idea is to match the exact continuum operator solution to the solution obtained after projecting onto the discrete basis. Figure~\ref{fig:Gap-vs-N-freestream} compares these analytic approximations to the numerical results obtained in this work.

We start from the similarity-transformed Hamiltonian given in Eq.~\eqref{eq:Htilde}.
Its diagonal elements are $\widetilde H_{l,l}=A_l^m$, while its
off-diagonal elements become (recall $l = l_0 + 2n$)
\begin{align}
    t_l
    &\equiv
    \sqrt{\frac{l-1}{l+4}}\,B_{l+2}^m
    =\widetilde H_{l+2,l}
    =-\widetilde H_{l,l+2}
    \notag\\
    &=
    \frac{l}{4}+\frac{3}{8}
    -\frac{8m^2+23}{32l}
    +O(l^{-2})\notag\\
    &=
    \frac{n}{2}+\frac{l_0}{4}+\frac{3}{8}
    -\frac{8m^2+23}{64n}
    +O(n^{-2}).
    \label{eq:t-asymptotic-n}
\end{align}
The diagonal coefficient can be written exactly as
\begin{align}
    A_l^m
    =
    \frac{7}{4}
    -\frac{5(4m^2-1)}{4(2l-1)(2l+3)}
    =
    \frac{7}{4}+O(l^{-2}).
    \label{eq:A-exact-asymptotic}
\end{align}
Therefore, \begin{align} \widetilde H = \frac{7}{4}I+K+V, \qquad K^\dagger=-K, \qquad V_{ll}=O(l^{-2}). \end{align} At large $l$, the diagonal correction $V$ vanishes, while the anti-Hermitian operator $K$ contributes only to the imaginary part of the eigenvalue. The large-$N$ solutions constructed below contain all frequencies $\omega\in\mathbb R$. Other than isolated eigenvalues, the continous spectrum of $\widetilde{H}$ therefore lies on the vertical line $\lambda =  \frac{7}{4}+i\mathbb R$.

We next determine the continuum modes and their connection coefficients from the original differential operator. In the
angular variable $x=\cos\theta$, the free-streaming operator is
\begin{align}
    H_F
    =
    4x^2-x(1-x^2)\partial_x.
    \label{eq:HF-coordinate}
\end{align}
Away from $x=0$ and $|x|=1$, the equation
$H_Ff_\lambda=\lambda f_\lambda$ has the parity-even and parity-odd
solutions
\begin{align}
\begin{split}    
    f_\lambda^{(+)}(x)
    &=
    |x|^{-\lambda}(1-x^2)^{(\lambda-4)/2},
    \\
    f_\lambda^{(-)}(x)
    &=
    \operatorname{sgn}(x)|x|^{-\lambda}
    (1-x^2)^{(\lambda-4)/2}.
    \label{eq:coordinate-generalized-solutions}
\end{split}
\end{align}

Next, we project the continuum solutions \eqref{eq:coordinate-generalized-solutions} onto the discrete basis of associated Legendre polynomials used in the Hamiltonian $H_\text{free}$ of \eqref{eq:H-free-app}. Using the same bracket defined in Eq.~\eqref{eq:inner-product} and the associated Legendre polynomials $\mathcal{P}_l^m(x) = e^{-im\phi} Y_l^m(\theta, \phi)$, we obtain the projected moments
\begin{align}
    q_l^{(s)}(\lambda)
    &=
    \left\langle \mathcal P_l^m,f_\lambda^{(s)}\right\rangle,
    \qquad
    \widetilde q_l^{(s)}=\sqrt{p_l}\,q_l^{(s)}.
\end{align}

At large $l$, $\mathcal P_l^m$ is highly oscillatory, such that after averaging $\left\langle \mathcal P_l^m,f_\lambda^{(s)}\right\rangle$ the leading terms come from the singular regions $x\sim0$ and
$|x|\sim1$:
\begin{align}
    \widetilde q_l^{(s)}(\lambda)
    \sim
    \left.\widetilde q_l^{(s)}(\lambda)\right|_{x\sim0}
    +
    \left.\widetilde q_l^{(s)}(\lambda)\right|_{|x|\sim1}.
\end{align}

For fixed $m$ and $l\to\infty$, the $x\rightarrow 1$ (or equivalently $0<\theta\ll 1$) limit is
\begin{align}
    \mathcal P_l^m(\cos\theta)
    =
    (-1)^m(2\nu_l)^{1/2}
    \left[J_m(\nu_l\theta)+O(\nu_l^{-1})\right],
    \qquad
    \theta=O(\nu_l^{-1}),
    \label{eq:legendre-endpoint-limit}
\end{align}
with $\nu_l=l+\frac12$. Using $\sqrt{p_l}\sim\nu_l^{-5/4}$ and
$\widetilde\theta=\nu_l\theta$, the endpoint contribution is
proportional to
\begin{align}
    (-1)^m\sqrt{2}\nu_l^{5/4-\lambda}
    \int_0^\infty d\widetilde\theta\,
    \widetilde\theta^{\lambda-3}J_m(\widetilde\theta) =
    (-1)^m2^{\lambda-5/2}
    \frac{
        \Gamma\left(\frac{m+\lambda-2}{2}\right)
    }{
        \Gamma\left(\frac{m-\lambda+4}{2}\right)
    }\times \nu_l^{5/4-\lambda}\,.\label{eq:endpoint-integral}
\end{align}
This gives the non-alternating power-law solution $q_l \sim \nu_l^{5/4-\lambda}$.

Near $x=0$, the two parity sectors satisfy
\begin{align}
    \mathcal P_{m+2n}^m(x)
    &=
    (-1)^{m+n}\sqrt{\frac{4}{\pi}}
    \left[
        \cos(\nu_{m+2n}x)+O(\nu_{m+2n}^{-1})
    \right],
    \\
    \mathcal P_{m+1+2n}^m(x)
    &=
    (-1)^{m+n}\sqrt{\frac{4}{\pi}}
    \left[
        \sin(\nu_{m+1+2n}x)+O(\nu_{m+1+2n}^{-1})
    \right].
\end{align}
With $\widetilde x=\nu_lx$, the corresponding contributions are
\begin{align}
\begin{split}
\label{eq:center-integrals}
    \left.\widetilde q_{m+2n}^{(+)}\right|_{x\sim0}
    &\propto
    (-1)^{m}\sqrt{\frac{4}{\pi}}
    \Gamma(1-\lambda)
    \sin\left(\frac{\pi\lambda}{2}\right) \times (-1)^n\nu_{m+2n}^{\lambda-9/4}, \\
    \left.\widetilde q_{m+1+2n}^{(-)}\right|_{x\sim0}
    &\propto
    (-1)^{m}\sqrt{\frac{4}{\pi}}
    \Gamma(1-\lambda)
    \cos\left(\frac{\pi\lambda}{2}\right) \times (-1)^n\nu_{m+1+2n}^{\lambda-9/4}.
\end{split}
\end{align}
They give the power-law solution with alternating sign in $n$. Both power laws also satisfy the discrete eigenvalue equation $\widetilde{H}\tilde{q}=\lambda\tilde{q}$,
\begin{align}
    t_{n-1} \tilde{q}_{l_{n-1}}
    +\left(A_{l_n}^m-\lambda\right) \tilde{q}_{l_{n}}
    -t_n  \tilde{q}_{l_{n+1}}
    =0
    \label{eq:eigenvector-recurrence}
\end{align}
up to relative $O(n^{-1})$ corrections. 

Because, in the $N\rightarrow \infty$ limit, we know that $\textrm{Re}\{\lambda\}=\frac{7}{4}$ other than for the isolated pole, we reparametrize $\lambda$ as
\begin{align}
    \lambda=\frac{7}{4}+\zeta.
\end{align}
The solution satisfying the lower boundary condition has the
large-$n$ form
\begin{align}
    \widetilde q_n^{(s)}(\zeta)
    &=\left\langle \mathcal P_l^m,f_{\frac{7}{4}+\zeta}^{(s)}\right\rangle\nonumber\\
    &=A_s(\zeta)\nu_{l_n}^{-1/2-\zeta}
    +(-1)^nB_s(\zeta)\nu_{l_n}^{-1/2+\zeta}
    +\cdots,
    \label{eq:connection-expansion}
\end{align}
where in the second line we have substituted the approximate solutions obtained by projecting onto the momentum-space basis in Eqs.~\eqref{eq:endpoint-integral} and \eqref{eq:center-integrals}.

The ratio of the two coefficients is
\begin{align}
    \frac{A_s(\zeta)}{B_s(\zeta)}
    =
    R_{m,s}\left(\frac{7}{4}+\zeta\right),
    \label{eq:def-connection-ratio}
\end{align}
where
\begin{align}
    R_{m,+}(\lambda)
    &=
    \sqrt{\frac{\pi}{4}}\,
    \frac{
        2^{\lambda-5/2}
        \Gamma\left(\frac{m+\lambda-2}{2}\right)
    }{
        \Gamma\left(\frac{m-\lambda+4}{2}\right)
        \Gamma(1-\lambda)
        \sin\left(\frac{\pi\lambda}{2}\right)
    },
    \label{eq:R-plus}
    \\
    R_{m,-}(\lambda)
    &=
    \sqrt{\frac{\pi}{4}}\,
    \frac{
        2^{\lambda-5/2}
        \Gamma\left(\frac{m+\lambda-2}{2}\right)
    }{
        \Gamma\left(\frac{m-\lambda+4}{2}\right)
        \Gamma(1-\lambda)
        \cos\left(\frac{\pi\lambda}{2}\right)
    }.
    \label{eq:R-minus}
\end{align}

At $\lambda=1$, $R_{m,+}(1)=0$ because $\Gamma(1-\lambda)\sin(\frac{\pi}{2}\lambda)$ blows up at $\lambda=1$, whereas $R_{m,-}(1)$ remains
finite and nonzero. In the even sector the coefficient of the
non-normalizable solution therefore vanishes, leaving
$\widetilde q_n^{(+)}\sim(-1)^n n^{-5/4}$ as a normalizable eigenvector.  Hence $\lambda=1$ is an
isolated even-sector eigenvalue, while the continuum has
$\Re(\lambda)=7/4$.  It follows immediately that
\begin{align}
    \Delta_\infty^{(+)}=\frac74-1= \frac34,
    \qquad
    \Delta_\infty^{(-)}=0.
\end{align}

For an $N$-moment truncation, $n=0,\ldots,N-1$, we can estimate the leading correction as follows. The first omitted
component must vanish, and Eq.~\eqref{eq:connection-expansion}
therefore gives
\begin{align}
    \widetilde q_N^{(s)}(\zeta)
    =
    A_s(\zeta)\nu_{l_N}^{-1/2-\zeta}
    +(-1)^NB_s(\zeta)\nu_{l_N}^{-1/2+\zeta}
    +\cdots
    =0,
    \label{eq:finite-N-boundary}
\end{align}
which is solved by
\begin{align}
    R_{m,s}\left(\frac74+\zeta\right)
    =
    (-1)^{N+1}\nu_{l_N}^{2\zeta}
    \left[1+O(N^{-1})\right].
    \label{eq:finite-N-quantization}
\end{align}
To obtain the leading correction, we take the real part of the logarithm and expand about $\zeta=0$. This gives the continuum edge
\begin{align}
    \Re\zeta_{\mathrm{edge},N}^{(+)}
    =
    \frac{
        \ln\left|R_{m,+}\left(\frac74\right)\right|
    }{
        2\ln\nu_{l_N}
    }
    +O\left((\ln N)^{-2}\right).
\end{align}
The isolated level satisfies $\lambda_{0,N}^{(+)}=1+O(N^{-3/2})$.
Since $\nu_{l_N}=m+2N+1/2$ in this sector, the even-sector gap is
therefore
\begin{align}
    \Delta_N^{(+)}
    =
    \frac34
    +\frac{
        \ln\left|R_{m,+}\left(\frac74\right)\right|
    }{
        2\ln\left(m+2N+\frac12\right)
    }
    +O\left((\ln N)^{-2}\right)
    +O(N^{-3/2}).
    \label{eq:even-gap}
\end{align}
This expression can also be used as the initial guess for a Newton iteration of Eq.~\eqref{eq:finite-N-quantization}.

Finally, consider the odd sector and odd $N$, for which
$(-1)^{N+1}=1$. 
The real branch solution $\zeta_{0,N}$ and the adjacent complex branch solution $\zeta_{1,N}$ are (writing $\mathcal L_N^{(-)}=\ln\left(m+2N+\frac32\right)$)
\begin{align}\label{eq:odd-branch-equations}
\begin{split}
    2\mathcal L_N^{(-)}\zeta_{0,N}
    &=
    \ln \left(R_{m,-}\left(\frac74+\zeta_{0,N}\right)\right),\\
    2\mathcal L_N^{(-)}\zeta_{1,N}
    &=
    \ln \left(R_{m,-}\left(\frac74+\zeta_{1,N}\right)\right)+2\pi i.
\end{split}
\end{align}
A convenient initial guess for solving
Eq.~\eqref{eq:odd-branch-equations} is obtained by replacing the $\zeta$ on the right-hand side by $\zeta=0$. The two initial guesses can then be improved by Newton iteration. Since $\mathcal{L}_N^{(-)}\to \infty$ as $N \to \infty$, we must have $\zeta \to 0$, and we thus recover
\begin{align}
    \Delta_{N}^{(-)}
    =
    \Re\left(\zeta_{1,N}-\zeta_{0,N}\right) \to 0 = \Delta^{(-)}_\infty,
    \label{eq:odd-gap}
\end{align}
for odd $N$. For even $N$, the two closest branches form a complex-conjugate pair and therefore have identical real parts. Figure~\ref{fig:Gap-vs-N-freestream} compares the numerical gaps and the asymptotic formulas and their one-step Newton improvements. Their convergence to $\Delta_\infty^{(+)}=3/4$ and $\Delta_\infty^{(-)}=0$ remains logarithmically slow.

\end{document}